# Interior solution of azimuthally symmetric case of Laplace equation in orthogonal similar oblate spheroidal coordinates

Pavel Strunz[a]

Nuclear Physics Institute of the CAS, 25068 Řež, Czech Republic

**Abstract** Curvilinear coordinate systems distinct from the rectangular Cartesian coordinate system are particularly valuable in the field calculations as they facilitate the expression of boundary conditions of differential equations in a reasonably simple way when the coordinate surfaces fit the physical boundaries of the problem. The recently finalized orthogonal similar oblate spheroidal (SOS) coordinate system can be particularly useful for a physical processes description inside or in the vicinity of the bodies or particles with the geometry of an oblate spheroid. The solution of the azimuthally symmetric case of the Laplace equation was found for the interior space in the orthogonal SOS coordinates. In the frame of the derivation of the harmonic functions, the Laplace equation was separated by a special separation procedure. A generalized Legendre equation was introduced as the equation for the angular part of the separated Laplace equation. The harmonic functions were determined as relations involving generalized Legendre functions of the first and of the second kind. Several lower-degree functions are reported. Recursion formula facilitating determination of the higher-degree harmonic functions was found. The general solution of the azimuthally symmetric Laplace equation for the interior space in the SOS coordinates is reported.

## 1 Introduction

Curvilinear coordinate systems different from the rectangular Cartesian coordinate system are of value in simple geometric considerations as the cumbersome mathematical expressions for determination, for example, of areas and volumes can be overcome. Moreover, curvilinear coordinate systems are particularly valuable in the field theory and differential equations solution. In order to express boundary conditions of differential equations in a reasonably simple way, coordinate surfaces that fit the physical boundaries of the problem are essential [1]. Therefore, a range of field problems that can be handled effectively depend upon the number of well-developed coordinate systems. Particularly, the orthogonal curvilinear coordinate systems [2] appear to be the most useful.

The recently finalized similar oblate spheroidal (SOS) orthogonal coordinate system [3, 4] can be a powerful tool for a description of physical processes inside or in the vicinity of the bodies with geometry of an oblate spheroid. Such bodies range (but are not limited to) from planets with a small oblateness (like the Earth with the ratio of the semi-axes difference to the major semi-axis length ≈1:300), through elliptical galaxies up to significantly flattened objects like disk galaxies.

In the field of atmospheric physics, it was stated [3] that the SOS coordinates could be of help for better modeling of geopotential surfaces, allowing for a better description of the spatial variation of the apparent gravity. Still more advantageously, the SOS coordinates could be used for modeling of the potential and atmosphere of significantly more oblate celestial bodies, e.g., the gas giants. Further, similar oblate spheroids are frequently used for modeling of iso-density levels inside galaxies [5, 6]. Therefore, the SOS coordinates can be helpful also in this field.

In electrostatics and solid-state physics, the SOS coordinates could find application in description of electric field potential in ferroelectric materials (e.g., ferroelectric nanocomposites) containing dielectric inclusions (or vice versa—ferroelectric inclusions in dielectric matrix) [7], particularly the inclusions of spheroidal shape [8].

The similar oblate spheroidal coordinates are distinct from all the standard orthogonal coordinate systems [2, 9] including the confocal oblate spheroidal system, as one coordinate surfaces family of the SOS coordinates is not of the second degree (and even not of the fourth degree) but it is formed by general power functions $z \approx x^{1+\mu}$ (i.e., with a real-number exponent $1 + \mu$) rotated around the $z$-axis [3, 4]. They are orthogonal to the similar oblate spheroids representing the first set of the coordinate surfaces. (For terminological clarity, a spheroid means in this text an ellipsoid of revolution or rotational ellipsoid. An oblate spheroid is a quadric surface obtained by rotating an ellipse about the shorter principal axis).

---

[a] e-mail: strunz@ujf.cas.cz (corresponding author)

Already reported [4] were the analytical coordinate transformation from the SOS coordinates to the Cartesian system as well as the metric scale factors and the Jacobian determinant. Further, the formulas necessary for the transformation of a vector field between the SOS system and the Cartesian coordinates were recently published [10]. Generalized sine and cosine applicable in the transformation were introduced as well. The derived analytical relations cannot be expressed in a closed form; nevertheless they employ convergent infinite power series (with generalized binomial coefficients) and are thus still analytical.

Nevertheless, there are still missing significant parts of algebra connected with the SOS coordinate system. Predominantly, solutions of standard differential equations in the SOS coordinates are not derived. It is known that several important scalar differential equations can be reduced to the Helmholtz equation or to its special case, the Laplace equation [2]. Therefore, the scalar fields in several areas of physics can be based on the solutions of the Helmholtz and of the Laplace equations.

This article deals with the task to solve the simpler of the two types of the abovementioned differential equations, the Laplace equation, in the SOS coordinate system. A solution of the Laplace equation would represent an important step in the development of the SOS coordinate system and its applicability in astrophysics (e.g., for modeling of gravitational potential) and in other fields of physics (cosmology, geodesy, climatology, electromagnetism). Determination of harmonic functions could also help to find solutions of the other, more complex, differential equations in the SOS coordinates.

The organization of the text is following. First, a short summary of the SOS coordinate system is provided. Then, the azimuthally symmetric Laplace equation in the SOS coordinates is formulated and a possible shape of the harmonic function is modeled. Special separation procedure is reported next, and the radial part of the equation is solved. Simple but informative enough solutions of the angular part of the separated equation are found. The found simple solutions as well as the Laplace equation itself are rewritten to the form enabling solution generalization. A new type of polynomials—generalizing the Legendre ones—are found to be a part of the harmonic functions in the SOS coordinates for the interior space. A recursion formula for the polynomials is derived. Finally, the complete interior solution of the Laplace equation in the azimuthally symmetric case in the SOS system is reported.

Occasionally during the derivation, the found interim results for the limiting case of oblate spheroids, i.e., for the spherical case, are compared with the well-known solution for the spherical coordinates.

A number of formulas needed within the article and derived on the basis of binomial identities are reported in Appendix A. Nevertheless, due to the fact that some derivations with an extensive use of combinatorial identities are very lengthy when written in a full detail, a large part of them are moved to Supplements to this article.

## 2 The SOS coordinates

The SOS coordinates were introduced by White et al. [3]. The full description of the analytical solution of SOS coordinates can be found in [4]. A shortened, but more overviewable summary is reported in [10]. It is recommended to find details of the system in the abovementioned references. A limited description (to the extent needed for the tasks of this article) is given also in this section.

For the SOS coordinate system $(R, \nu, \lambda)$ [4], the basic coordinate surfaces of the $R$ coordinate are similar oblate spheroids given in 3D Cartesian coordinates $(x_{3D}, y_{3D}, z_{3D})$ by the formula

$$x_{3D}^2 + y_{3D}^2 + (1+\mu)z_{3D}^2 = R^2 \tag{1}$$

The $R$ coordinate value is equal to the equatorial radius of the particular spheroid from the family. The parameter $\mu$ characterizes the whole family of the similar oblate spheroids. The parameter $\mu > 0$ for oblate spheroids, and the minor and major semi-axes of each member of the spheroid family have the ratio $(1+\mu)^{-1/2}$. As a limit (when $\mu = 0$), a sphere (and spherical coordinate surfaces) is determined by (1). A special reference spheroid is introduced with the equatorial radius $R_0$, usually coinciding with the reference surface of the object for which the SOS coordinate system is to be applied.

The second set of the coordinate surfaces, orthogonal to the similar oblate spheroids defined above, are power functions in 3D of the shape [3, 4]

$$z_{3D} \sim \left(\sqrt{x_{3D}^2 + y_{3D}^2}\right)^{1+\mu} \tag{2}$$

The labeling, i.e., the coordinate $\nu$ corresponding to these surfaces, is equivalent to the so-called parametric latitude [4]. The coordinate $\nu$ is also equivalent to the parameter used for the standard parametric equation of the ellipse, representing the meridional section of the reference spheroid mentioned above, having a special equatorial radius (major semi-axis) equal to $R_0$, i.e.,

$$\sqrt{x_{3D}^2 + y_{3D}^2} = R_0 \cos \nu \quad \text{and} \quad z_{3D} = \frac{R_0}{\sqrt{1+\mu}} \sin \nu \tag{3}$$

Using this definition, the coordinate $\nu$ is equal to zero in the equatorial plane while it is equal to $\pm \pi/2$ on the rotation axis (+ in the north,—in the south). From Eqs. (3) and (17) of [4], the relation for the power function coordinate surfaces in 3D (2) in dependence on the coordinate $\nu$ can be detailed as

$$z_{3D} = \frac{1}{\sqrt{1+\mu}} \frac{1}{R_0^\mu} \frac{\sin \nu}{\cos^{1+\mu} \nu} \left(\sqrt{x_{3D}^2 + y_{3D}^2}\right)^{1+\mu} \tag{4}$$

(the correctness of the relation can be easily tested when cos∘$\nu$ and sin∘$\nu$ according to (3) are plugged into this relation).

Finally, the third set of the coordinate surfaces, orthogonal to the previous two, are semi-infinite planes containing the rotation axis. The associated coordinate is the longitude angle $\lambda$, which is the same as its equivalent coordinate in the spherical coordinate system.

A key role in the derivation of the SOS coordinate algebra (e.g., the coordinates transformation between the SOS and the Cartesian system [4]) plays the dimensionless parameter $W$ defined as

$$W = \left(\frac{R}{R_0}\right)^\mu \frac{\sin \nu}{\cos^{1+\mu} \nu} \tag{5}$$

which will be used frequently in this article. The constant-$W$ surfaces are straight half-lines starting at the origin and rotated around the system axis, i.e., cones [4]. In what follows, the calculation is restricted to the parametric latitude $\nu \in 0, \frac{\pi}{2}$. With this limitation, the parameter $W$, see Eq. (5), is always nonnegative which simplifies further derivations. Due to the symmetry (reflection with respect to the equator), expressions and solutions relevant for the SOS system in the complementary range $\nu \in -\frac{\pi}{2}, 0$) can be easily obtained.

The SOS coordinates can be transformed to the Cartesian coordinates using analytical expressions including infinite power series with generalized binomial coefficients [4, 10]. Relations enabling such approach were first reported by Pólya and Szegö [11]. Due to the convergency limits of the involved power series, the expressions were derived separately in so-called "the small-$\nu$ region" and separately in "the large-$\nu$ region" [4]. The border between the two regions is defined by the parameter $W$ value (see Eq. (5)) fulfilling the following relation:

$$W_{\text{border}}(R, \nu) = \sqrt{\frac{\mu^\mu}{(1+\mu)^{1+\mu}}} = \text{constant} \tag{6}$$

This border surface is a straight half-line starting at the origin and rotated around the system axis (and forming thus a surface of a cone) [4]. Although $W_{\text{border}}$ is a constant for a fixed $\mu$, the coordinates $R$ and $\nu$ vary along the half-line.

## 3 Important basic relations of the SOS coordinate system

In order to fulfill the aim of this article, some already derived relations for the SOS coordinates [4, 10] are needed. Therefore, the relevant ones are listed in this section. The formulae are listed separately for the small-$\nu$ region and for the large-$\nu$ region.

First needed relations are the metric scale factors [4, 10]. Below are listed the metric scale factors for $R$ coordinate,

$$h_R = \sqrt{\sum_{k=0}^{\infty} \binom{-\mu k}{k} (W^2)^k} \quad \text{and} \quad h_R = \frac{1}{\sqrt{1+\mu}} \sqrt{\sum_{k=0}^{\infty} \binom{\frac{\mu}{1+\mu} k}{k} \left(W^{-\frac{2}{1+\mu}}\right)^k} \tag{7}$$

as well as for $\nu$ coordinate,

$$h_\nu = \frac{R}{\sqrt{1+\mu}} \frac{\partial W}{\partial \nu} \sqrt{\sum_{k=0}^{\infty} \binom{-(\mu+2)-\mu k}{k} (W^2)^k}$$

and

$$h_\nu = \frac{R}{1+\mu} W^{-\frac{2+\mu}{1+\mu}} \frac{\partial W}{\partial \nu} \sqrt{\sum_{k=0}^{\infty} \binom{-\frac{2+\mu}{1+\mu} + \frac{\mu}{1+\mu} k}{k} \left(W^{-\frac{2}{1+\mu}}\right)^k} \tag{8}$$

, respectively. The left side (or the first one) of the above relations shows the formula valid for the small-$\nu$ region while the right side shows the formula valid for the large-$\nu$ region. The partial derivative in the $h_\nu$ relations can be expressed as

$$\frac{\partial W}{\partial \nu} = \left(\frac{R}{R_0}\right)^\mu \frac{1 + \mu \sin^2 \nu}{\cos^{2+\mu} \nu} = \frac{1 + \mu \sin^2 \nu}{\sin \nu \cos \nu} W \tag{9}$$

For completeness, we also report the partial derivative of $W$ with respect to $R$, which is easily obtained from (5)

$$\frac{\partial W}{\partial R} = \frac{\mu}{R} \left(\frac{R}{R_0}\right)^\mu \frac{\sin \nu}{\cos^{1+\mu} \nu} = \frac{\mu}{R} W \tag{10}$$

Further, the already derived Jacobian determinant $\Im = h_R h_\nu h_\lambda$ [4, 10] is listed:

$$\Im = \frac{R^2}{\sqrt{1+\mu}} \frac{\partial W}{\partial \nu} \sum_{k=0}^{\infty} \binom{-\frac{\mu+3}{2} - \mu k}{k} (W^2)^k$$

$$\Im = \frac{R^2 W^{-\frac{3+\mu}{1+\mu}}}{\sqrt{(1+\mu)^3}} \frac{\partial W}{\partial \nu} \sum_{k=0}^{\infty} \binom{-\frac{1}{2}\frac{\mu+3}{1+\mu} + \frac{\mu}{1+\mu}k}{k} \left(W^{-\frac{2}{1+\mu}}\right)^k \tag{11}$$

Again, the first relation shows the analytic formula valid for the small-$\nu$ region, while the second relation shows the formula valid for the large-$\nu$ region. The Jacobian divided by the square of the $h_R$ metric scale factor is expressed as

$$\frac{\Im}{h_R^2} = \frac{R^2}{\sqrt{1+\mu}} \frac{\partial W}{\partial \nu} \sum_{k=0}^{\infty} \frac{-\frac{\mu+3}{2}}{-\frac{\mu+3}{2} - \mu k} \binom{-\frac{\mu+3}{2} - \mu k}{k} (W^2)^k \text{ and}$$

$$\frac{\Im}{h_R^2} = \frac{R^2}{\sqrt{1+\mu}} W^{-\frac{3+\mu}{1+\mu}} \frac{\partial W}{\partial \nu} \sum_{k=0}^{\infty} \frac{-\frac{1}{2}\frac{\mu+3}{\mu+1}}{-\frac{1}{2}\frac{\mu+3}{1+\mu} + \frac{\mu}{1+\mu}k} \binom{-\frac{1}{2}\frac{\mu+3}{1+\mu} + \frac{\mu}{1+\mu}k}{k} \left(W^{-\frac{2}{1+\mu}}\right)^k \tag{12}$$

for the small-$\nu$ region and for the large-$\nu$ region, respectively [4, 10].

The Jacobian divided by the square of the $h_\nu$ metric scale factor is derived in Appendix A of this article both for the small-$\nu$ region (the first relation, (A8) in the appendix) and for the large-$\nu$ region (the second relation, (A9) in the appendix):

$$\frac{\Im}{h_\nu^2} = \frac{\sqrt{1+\mu}}{\frac{\partial W}{\partial \nu}} \sum_{k=0}^{\infty} \frac{\frac{\mu+1}{2}}{\frac{\mu+1}{2} - \mu k} \binom{\frac{\mu+1}{2} - \mu k}{k} (W^2)^k \text{ and}$$

$$\frac{\Im}{h_\nu^2} = \frac{W\sqrt{1+\mu}}{\frac{\partial W}{\partial \nu}} \sum_{k=0}^{\infty} \frac{\frac{1}{2}}{\frac{1}{2} + \frac{\mu}{1+\mu}k} \binom{\frac{1}{2} + \frac{\mu}{1+\mu}k}{k} \left(W^{-\frac{2}{1+\mu}}\right)^k \tag{13}$$

Finally, the functions for the generalized cosine ($f_C$) and for the generalized sine ($f_S$) [10] in the frame of the SOS coordinate system are reported:

$$f_C = \sqrt{\sum_{k=0}^{\infty} \binom{-1 - \mu k}{k} (W^2)^k} \text{ and}$$

$$f_C = \frac{W^{-\frac{1}{1+\mu}}}{\sqrt{1+\mu}} \sqrt{\sum_{k=0}^{\infty} \binom{-\frac{1}{1+\mu} + \frac{\mu}{1+\mu}k}{k} \left(W^{-\frac{2}{1+\mu}}\right)^k} \tag{14}$$

$$f_S = W\sqrt{1+\mu} \sqrt{\sum_{k=0}^{\infty} \binom{-(1+\mu) - \mu k}{k} (W^2)^k} \text{ and } f_S = \sqrt{\sum_{k=0}^{\infty} \binom{-1 + \frac{\mu}{1+\mu}k}{k} \left(W^{-\frac{2}{1+\mu}}\right)^k} \tag{15}$$

The left sides (the first realtions) show the formulae valid for the small-$\nu$ region, while the right sides (the second relations) for the large-$\nu$ region. As $f_S, f_C$ depend solely on $W$ (not separately on $R$ and $\nu$), these functions are constant at the same surfaces in 3D where $W$ is constant. As (see the text after (5)) the constant-$W$ surfaces are cones with the tip at the origin and the axis coinciding with the rotation axis of the coordinate system, so are also $f_S, f_C$ constant on the same surfaces.

For $\mu = 0$, the generalized cosine and the generalized sine become cosine function and sine function, respectively. Similarly as for the standard cosine and sine, the following relation is valid for the generalized cosine and the generalized sine:

$$f_C^2 + f_S^2 = 1 \tag{16}$$

Further (see [10] or Appendix A, relation (A13), and using (16)) the relations of the generalized cosine and of the generalized sine to the metric scale factor $h_R$ are

$$f_S^2 = \frac{1+\mu}{\mu}(1 - h_R^2), \; f_C^2 = \frac{(1+\mu)h_R^2 - 1}{\mu} \tag{17}$$

The functions listed in this section all contain infinite power series with generalized binomial coefficients. Operations with them are not as straightforward as a simple multiplication or summation of two expressions. Nevertheless, it is possible to deal with them with a help of known combinatorial identities [11–15]. The important ones, used in the derivations also in this article, are listed in [10]. With their help, a number of more complex relations needed for the Laplace equation solution are derived in Appendix A and subsequently used in the following sections.

## 4 Laplace equation in SOS coordinates in azimuthally symmetric case

The Laplace equation,

$$\Delta V = 0 \tag{18}$$

in the SOS coordinates can be derived using the general formula for the Laplace operator $\Delta$. In general curvilinear orthogonal coordinates, it is reported in [16], Eq. (15). Applied to the SOS case and the symbols for the SOS coordinates $R$, $\nu$, $\lambda$, the operator can be written as

$$\Delta V = \frac{1}{\Im}\left[\frac{\partial}{\partial R}\left(\frac{\Im}{h_R^2}\frac{\partial V(R,\nu,\lambda)}{\partial R}\right) + \frac{\partial}{\partial \nu}\left(\frac{\Im}{h_\nu^2}\frac{\partial V(R,\nu,\lambda)}{\partial \nu}\right) + \frac{\partial}{\partial \lambda}\left(\frac{\Im}{h_\lambda^2}\frac{\partial V(R,\nu,\lambda)}{\partial \lambda}\right)\right] \quad (19)$$

where $V(R,\nu)$ denotes the potential function (or generally any function with physical meaning on which the Laplace operator is to be applied) and $\Im = h_R h_\nu h_\lambda$ is the Jacobian determinant.

The Laplace Eq. (18) in the SOS coordinates in azimuthally symmetric case (i.e., when the potential function is independent of $\lambda$) is then

$$\frac{\partial}{\partial R}\left(\frac{\Im}{h_R^2}\frac{\partial V(R,\nu)}{\partial R}\right) + \frac{\partial}{\partial \nu}\left(\frac{\Im}{h_\nu^2}\frac{\partial V(R,\nu)}{\partial \nu}\right) = 0 \quad (20)$$

The basic strategy for its solution is to modify the equation algebraically in such a way that it enables variables separation, and thus transformation of the partial differential equation to a set of ordinary differential equations.

It follows from the shape of the involved functions (the metric scale factors and the Jacobian) in the SOS case that a standard separation procedure using Stäckel matrix (see e.g., [2]) is not applicable. Therefore, either separation is not possible, or it is a non-standard one, and another way for separation thus has to be searched for.

To test which of the abovementioned options is correct, a model potential was input into the Laplace Eq. (20). With the experience with the Laplace equation in the spherical coordinates, the model of the solution of the Laplace equation in the similar oblate spheroidal coordinates was first tested in the form of a product of three functions plus a constant $V_C$:

$$V_1(R,\nu) = R^\alpha F(W) G(\nu) + V_C \quad (21)$$

It was found in the first round of derivation that at least one (but certainly not the only one) solution in this form exists, and it is equal to

$$V_1(R,\nu) = c_1 R \frac{f_S(W)}{h_R(W)} + V_C \quad (22)$$

where $c_1$ is an arbitrary real constant, $f_S$ is the generalized sine (15) and $h_R$ is the metric scale factor of the $R$ coordinate in the SOS coordinate system [4]. Both functions $f_S$ and $h_R$ depend on the composed variable $W$ (see Eq. (5)), not individually on the coordinates $R$ and/or $\nu$. Although the derivation of the result (22) is non-trivial, it is not reported here in detail as the sequence of the steps leading to it is very similar to the one which will be reported in the second applied model further in the text.

The found result (22) indicates that—at least for some solutions of the Laplace equation in the SOS coordinates, or for some region in the space—the solution can be written without an explicitly expressed dependence on $\nu$ coordinate, only with the explicit dependence on $R$ and on $W$, moreover appearing in two separate functions, a product of which (plus a constant) represents the solution. Therefore, the second applied model function reflecting this finding and to be tested was of the form

$$V_2(R,\nu) = r(R) F(W) + V_C \quad (23)$$

where $r$ and $F$ are functions of $R$ and $W$, respectively. This new model excludes $G(\nu)$ function contained in the model (21); on the other side, it tests a more general function $r(R)$ for the radial part of the model.

When we insert the model potential (23) into the Laplace equation in the azimuthally symmetric case (20), we obtain the equation in the form

$$\frac{\partial}{\partial R}\left(\frac{\Im}{h_R^2}\frac{\partial}{\partial R}(r(R)F(W))\right) + \frac{\partial}{\partial \nu}\left(\frac{\Im}{h_\nu^2}\frac{\partial}{\partial \nu}(r(R)F(W))\right) = 0 \quad (24)$$

By an intensive use of the product and of the chain rules, and by the use of relations (A96), (A97) and (A38) derived in Appendix A, the equation can be rewritten into the form

$$\left\{\left[(\mu-2)\mu + (\mu+3)\mu h_R^2 + \frac{h_R^4(1+\mu)^2}{f_C^2}\right]\frac{W}{F(W)}\frac{dF(W)}{dW} + \left[\mu^2 + \frac{h_R^4(1+\mu)^2}{f_C^2 f_S^2}\right]\frac{W^2}{F(W)}\frac{d^2F(W)}{dW^2}\right\} \\ + \left\{\frac{R^2}{r(R)}\frac{d^2r(R)}{dR^2} - \frac{R}{r(R)}\frac{dr(R)}{dR}\right\} + \left\{\left[(\mu+3)h_R^2 + 2\mu\frac{W}{F(W)}\frac{dF(W)}{dW}\right]\frac{R}{r(R)}\frac{dr(R)}{dR}\right\} = 0 \quad (25)$$

where $f_C$ and $f_S$ are the generalized cosine and sine reported in the previous chapter. This extensive derivation can be found in Supplement B, Sect. B1.

It can be easily checked that when setting $\mu = 0$ (i.e., in the spherical case) the equation simplifies (thanks to the characteristics of the generalized cosine and sine, as well as the metric scale factor $h_R$) to

$$\frac{1}{\cos^2 v} \frac{W}{F(W)} \frac{dF(W)}{dW} + \frac{1}{\cos^2 v \sin^2 v} \frac{W^2}{F(W)} \frac{d^2 F(W)}{dW^2}$$
$$+ \frac{R^2}{r(R)} \frac{d^2 r(R)}{dR^2} + 2 \frac{R}{r(R)} \frac{dr(R)}{dR} = 0 \qquad (26)$$

Further taking into account the form of derivatives of $F(W)$, see Eqs. (A44) and (A47) in Appendix A, we obtain—in this special case, i.e., for $\mu = 0$—the Laplace equation in the spherical coordinates:

$$\frac{1}{F(W)} \frac{\partial^2 F(W)}{\partial v^2} - \frac{\sin v}{\cos v} \frac{1}{F(W)} \frac{\partial F(W)}{\partial v} + \frac{R^2}{r(R)} \frac{d^2 r(R)}{dR^2} + 2 \frac{R}{r(R)} \frac{dr(R)}{dR} = 0 \qquad (27)$$

(it has to be taken into account that $v$ is latitude, not the polar angle, and that $W$, according to its definition by (5), does not depend on $R$ in the case when $\mu = 0$, but solely on $v$).

## 5 Separation of Laplace equation in the SOS coordinates

For a solution of Eq. (25), i.e., the Laplace equation in the SOS coordinates, we can apply a special variable separation approach. It can be seen that the first two terms in (25), containing the first and the second derivative of $F(W)$ and closed by curl brackets, depend only on the composed variable $W$, not explicitly on $R$, while the third and the fourth terms (the middle ones, containing the first and the second derivative of $r(R)$) depend only on $R$. The last term (also closed by curl brackets) is a product of a pure $W$-dependent function and a pure $R$-dependent function. It does worth to notice that there is no explicit dependence on the coordinate $v$, as also the functions $h_R$, $f_C$ and $f_S$ depend only on the composed variable $W$. The coordinate $v$ enters the equation only through $W$.

Equation (25) can be thus written in the following simplified form:

$$a(W) + b(R) + c(W)d(R) = 0 \qquad (28)$$

with a clear correspondence of the individual terms in (28) with the terms in (25). In order this equation being fulfilled for any $W$ and for any $R$, either $b(R)$ and $d(R)$ has to be equal to constants for any $R$, or $a(W)$ and $c(W)$ has to be equal to constants for any $W$. Otherwise, two different equations would arise for $a(W)$, $c(W)$ or for $b(R)$, $d(R)$ functions when keeping $R$ equal to a particular value $R_1$, and when keeping $R$ equal to another particular value $R_2$ (or also similarly when keeping $W = W_1$, and when keeping $W = W_2$). If $b(R_1)$ is not equal to $b(R_2)$, then the equations

$$a(W) + b(R_1) + c(W)d(R_1) = 0 \qquad (29)$$

$$a(W) + b(R_2) + c(W)d(R_2) = 0 \qquad (30)$$

would be generally two different equations for $a(W)$ and $c(W)$, and they could not be fulfilled at once for all possible $W$. It cannot be compensated by different values $d(R_1)$ and $d(R_2)$. The same reasoning can be done for other possible cases.

When the first possibility—i.e., $a(W)$ and $c(W)$ (according to (25)) are constants—is tested, it is concluded (after extensive algebraic operations not reported here) that the Laplace equation cannot be fulfilled in this case.

The second possibility—i.e., the case when $b(R)$ and $d(R)$ are constants—leads to the following relations (easily obtained by comparison of (28) and (25)):

$$d(R) = \frac{R}{r(R)} \frac{\partial r(R)}{\partial R} = K_d \qquad (31)$$

$$b(R) = \frac{R^2}{r(R)} \frac{\partial^2 r(R)}{\partial R^2} - \frac{R}{r(R)} \frac{\partial r(R)}{\partial R} = K_b \qquad (32)$$

where $K_d$, $K_b$ are separation constants. The first differential Eq. (31) has the solution for $r(R)$ in the form

$$r(R) = C_d R^{K_d} \qquad (33)$$

This relation tells us that the radial part of the Laplace equation (at least in some solution of the Laplace equation, or in some region in space) is basically the same as for the spherical case solution, except the equatorial radius $R$ is used instead of radius.

It follows from the second Eq. (32), see Supplement B, Sect. B2 for details, that the following equation connects the separation constants $K_b$ and $K_d$:

$$K_b = K_d(K_d - 2) \qquad (34)$$

Then, the remaining part of the Laplace Eq. (25), i.e., (in the abbreviated notation)

$$a(W) + K_d(K_d - 2) + c(W)K_d = 0 \tag{35}$$

can be written—with a help of the separation constant $K_d$—in the form (see (B36) in Supplement B, Sect. B2)

$$\left[\mu^2 + \frac{h_R^4(1+\mu)^2}{f_C^2 f_S^2}\right] \frac{W^2}{F(W)} \frac{d^2 F(W)}{dW^2}$$
$$+ \left[(\mu - 2)\mu + (\mu + 3)\mu h_R^2 + \frac{h_R^4(1+\mu)^2}{f_C^2} + 2K_d\mu\right] \frac{W}{F(W)} \frac{dF(W)}{dW} \tag{36}$$
$$+ K_d(\mu + 3)h_R^2 + K_d(K_d - 2) = 0$$

Although it is not a fully rigorous terminology, this remaining part of the Laplace equation (and its algebraic modifications) in the SOS coordinates—depending solely on the composed parameter $W$—will be sometimes denoted as "angular part of the Laplace equation" in the following text, as it is the part of the Laplace equation depending also on the meridional coordinate $v$ hidden in $W$.

## 6 Relatively simple but informative solution of the angular part of the Laplace equation

The further strategy for the solution is to modify algebraically (36) to the form containing only the metric scale factor $h_R$, not $f_C$ and $f_S$, and to see how that form can be solved. With the help of relations (17), and also (A22) from Appendix A, and after rather extensive algebraic manipulations (see Supplement B, Sect. B2 for details), (36) is modified to the form containing—in the pre-factors of the derivatives and of the function $F(W)$ itself—only powers of $h_R$ function and the parameter $W$:

$$\frac{d^2 F(W)}{dW^2} + \frac{1}{\mu W}$$
$$\frac{[(\mu^2 - \mu - 7) + 2K_d(2+\mu)]h_R^2 + [(\mu^2 + 8\mu + 9) - 2K_d(1+\mu)]h_R^4 + (-2\mu - 4)(1+\mu)h_R^6 + [2 - \mu - 2K_d]}{(2+\mu)h_R^2 - 1}$$
$$\frac{dF(W)}{dW} + \frac{K_d}{\mu^2 W^2} \tag{37}$$
$$\frac{[-3\mu - 7 + (2+\mu)K_d]h_R^2 + [(7\mu + 8 + \mu^2) - (1+\mu)K_d]h_R^4 - (1+\mu)(\mu + 3)h_R^6 - [K_d - 2]}{(2+\mu)h_R^2 - 1} F(W) = 0$$

Equation (37) is a rather complicated ordinary second order differential equation for $F(W)$, moreover depending also on the separation constant $K_d$. Its solution is not trivial and—before finding a general solution—it would be of advantage to find at least one relatively simple solution. There is certainly a trivial solution ($F(W)$ equal to zero) and—for $K_d = 0$ which makes the last term zero—another clear solution is also when $F(W)$ is a constant. However, these solutions are not informative enough to enable steps forward in finding all the possible solutions. Therefore, still relatively simple, but a non-trivial non-constant solution has to be searched for.

Equation (37) can be possibly written also with a help of an unknown $g(W)$ function as follows [17]:

$$\frac{\partial^2 F(W)}{\partial W^2} + f(W)\frac{\partial F(W)}{\partial W} + \left[f(W)g(W) - [g(W)]^2 + \frac{\partial g(W)}{\partial W}\right] F(W) = 0 \tag{38}$$

where $f(W)$ is the factor by the first derivative of $F(W)$ in (37), i.e.,

$$f(W) = \frac{1}{\mu W} \frac{[(\mu^2 - \mu - 7) + 2K_d(2+\mu)]h_R^2 + [(\mu^2 + 8\mu + 9) - 2K_d(1+\mu)]h_R^4 + (-2\mu - 4)(1+\mu)h_R^6 + [2 - \mu - 2K_d]}{(2+\mu)h_R^2 - 1} \tag{39}$$

By a comparison of the factor by the function $F(W)$ in (37) and (38), and by substituting $f(W)$ according to (39), we obtain the first order differential equation for $g(W)$ in the form:

$$\frac{K_d}{\mu^2 W^2} \frac{[-3\mu - 7 + (2+\mu)K_d]h_R^2 + [(7\mu + 8 + \mu^2) - (1+\mu)K_d]h_R^4 - (1+\mu)(\mu + 3)h_R^6 - [K_d - 2]}{(2+\mu)h_R^2 - 1}$$
$$= +\frac{1}{\mu W} \frac{[(\mu^2 - \mu - 7) + 2K_d(2+\mu)]h_R^2 + [(\mu^2 + 8\mu + 9) - 2K_d(1+\mu)]h_R^4 + (-2\mu - 4)(1+\mu)h_R^6 + [2 - \mu - 2K_d]}{(2+\mu)h_R^2 - 1} g(W)$$
$$- [g(W)]^2 + \frac{\partial g(W)}{\partial W} \tag{40}$$

Then, a solution of the second order differential equation for $F(W)$ is transferred to a solution of the first order differential equation for $g(W)$. If such $g(W)$ could be found, then an exact solution of the partial Laplace Eq. (37) would be possible to find as well (see [17]). Nevertheless, the above first order nonlinear differential Eq. (40) for $g(W)$, which is the Ricatti general equation [17], has to be solved first for $g(W)$.

Note that writing the Laplace equation (the separated part of it depending only on $W$, to be precise) in the special form of (38) equation is possibly not the only option how to find a solution. Probably, other solutions could be found in cases when the equation cannot be written in the form of (38). Nevertheless, (38) could lead to one of the simplest solutions. The search for another/others would be facilitated when at least one solution is known.

Further, we still simplify the task by setting $K_d = 1$, which could lead to still more specific solution of (40). Nevertheless, even in such case the equation is not trivial to solve. Again, extensive trials and algebraic derivations (see Supplement B, Sect. B3 for details), which use (A75), (A111), (A29) derived in Appendix A, are employed for solving (40). The particular solution for $K_d = 1$ is then found with the shape

$$g(W) = \frac{1}{\mu W}\left[1 - (1+\mu)h_R^2\right] = -\frac{f_C^2}{W} \tag{41}$$

where the generalized cosine function $f_C$ is given by (14). Equation (38), in which we know the function $g(W)$, has then the particular solution (see [17]):

$$F_0(W) = \exp\{-\int g(W)dW\} = \exp\left\{-\int \frac{1}{\mu W}[1 - (1+\mu)h_R^2]dW\right\} = \exp\left\{\frac{\mu+1}{\mu}\int \frac{h_R^2}{W}dW - \frac{1}{\mu}\int \frac{1}{W}dW\right\} \tag{42}$$

According to Appendix A, relation (A111), the integral containing the metric scale factor $h_R$ can be expressed in the following form:

$$\int W^{-1}h_R^2 dW = \ln\left(\frac{f_S}{f_C}\right) \tag{43}$$

and the particular solution of the angular part of the Laplace equation is thus

$$F_0(W) = \exp\left[\frac{\mu+1}{\mu}\ln\left(\frac{f_S}{f_C}\right) - \frac{1}{\mu}\ln W\right] = W^{-\frac{1}{\mu}}\left(\frac{f_S}{f_C}\right)^{\frac{\mu+1}{\mu}} \tag{44}$$

Thanks to the identities of [11], see Appendix A, (A1), (A2), (A3), (A4), (A5), this relation can be transformed to a still simpler expression (the derivation shown in Supplement B, Sect. B3), particularly

$$F_0(W) \sim \frac{f_S}{h_R} \tag{45}$$

where $f_S/h_R$ is the generalized sine divided by the metric scale factor of the coordinate $R$. According to (7), (14), (15), and also (A7) from Appendix A, we can write for $f_S/h_R$ the analytical expressions

$$\frac{f_S}{h_R} = W\sqrt{1+\mu}\sqrt{\frac{\sum_{k=0}^{\infty}\binom{-(1+\mu)-\mu k}{k}(W^2)^k}{\sum_{k=0}^{\infty}\binom{-\mu k}{k}(W^2)^k}} = W\sqrt{1+\mu}\sqrt{\sum_{k=0}^{\infty}\frac{-(1+\mu)}{-(1+\mu)-\mu k}\binom{-(1+\mu)-\mu k}{k}(W^2)^k} \tag{46}$$

in the small-$\nu$ region and

$$\frac{f_S}{h_R} = \sqrt{1+\mu}\sqrt{\frac{\sum_{k=0}^{\infty}\binom{-1+\frac{\mu}{1+\mu}k}{k}\left(W^{-\frac{2}{1+\mu}}\right)^k}{\sum_{k=0}^{\infty}\binom{\frac{\mu}{1+\mu}k}{k}\left(W^{-\frac{2}{1+\mu}}\right)^k}} = \sqrt{1+\mu}\sqrt{\sum_{k=0}^{\infty}\frac{-1}{-1+\frac{\mu}{1+\mu}k}\binom{-1+\frac{\mu}{1+\mu}k}{k}\left(W^{-\frac{2}{1+\mu}}\right)^k} \tag{47}$$

in the large-$\nu$ region. Notice that $f_S/h_R$ depends solely on $W$ (not separatelly on $R$ and $\nu$). Therefore, $f_S/h_R$ function is constant on the same surfaces in 3D where $W$ is constant, i.e., on the cones with the tip at the origin and the axis coinciding with the rotation axis of the coordinate system. For an axial 2D section through the coordinate system, these cones become straight half-lines radiating from the center. For a better imagination, $x$–$z$ plane map of the constant levels of $f_S/h_R$ function is shown in Fig. 1.

As $f_S/h_R$ appears to be a very important ratio in the Laplace equation solution, it does worth to derive some identities valid for this ratio. From the abovementioned derivation in Supplement B, Sect. B3, the following identity can be deduced:

$$\left(\sqrt{1+\mu}\right)^{\frac{1}{\mu}}\frac{f_S}{h_R} = W^{-\frac{1}{\mu}}\left(\frac{f_S}{f_C}\right)^{\frac{\mu+1}{\mu}} \Rightarrow \left(\frac{f_S}{h_R}\right)^{\mu} = \frac{1}{\sqrt{1+\mu}W}\left(\frac{f_S}{f_C}\right)^{1+\mu} \tag{48}$$

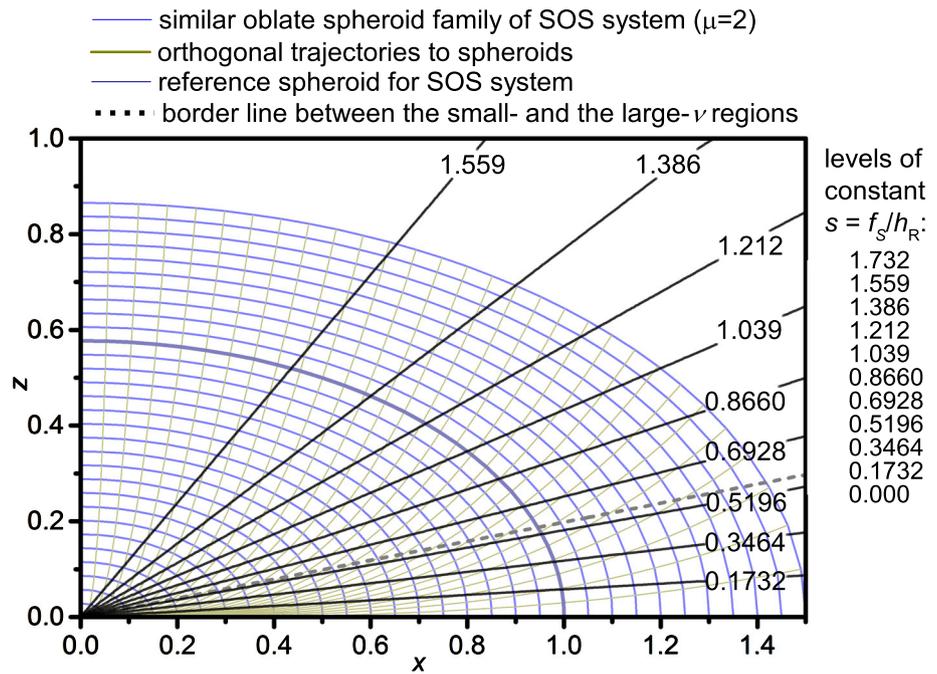

**Fig. 1** Constant levels of $f_S/h_R$ function (the generalized sine divided by the metric scale factor $h_R$) in one quadrant of $x$–$z$ plane for the SOS coordinate system with $\mu = 2$. The set of the orthogonal SOS coordinate system lines is used here as a background for a better overview

It can be further shown (see Eqs. (153) and (157) in [10]) that the ratio $f_S/h_R$ in a point is proportional to the ratio of the Cartesian coordinate $z_{3D}$ and the equatorial radius $R$ of the corresponding SOS coordinate spheroid surface for that point, i.e.,

$$\frac{f_S}{h_R} = (1+\mu)\frac{z_{3D}}{R} \Rightarrow z_{3D} = \frac{1}{1+\mu} R \frac{f_S}{h_R} \tag{49}$$

For the spherical case, where $\mu = 0$, $h_R = 1$, $f_S = \sin(\nu)$ and $R$ meaning the radius of a sphere, this relation becomes trivial: $\sin(\nu) = z_{3D}/R$.

We can also derive Cartesian coordinates $x_{3D}$, $y_{3D}$ in 3D as a function of $f_S/h_R$:

$$x_{3D}^2 = \cos^2\lambda R^2 \frac{f_C^2}{h_R^2} = \cos^2\lambda R^2 \left(1 - \frac{1}{1+\mu}\frac{f_S^2}{h_R^2}\right) \tag{50}$$

$$y_{3D}^2 = \sin^2\lambda R^2 \frac{f_C^2}{h_R^2} = \sin^2\lambda R^2 \left(1 - \frac{1}{1+\mu}\frac{f_S^2}{h_R^2}\right) \tag{51}$$

The derivation is done in Appendix A (see Eqs. (A25), (A26)). Therefore, the squared distance from the axis of rotation expressed in the SOS coordinate system is

$$x_{3D}^2 + y_{3D}^2 = R^2 \frac{f_C^2}{h_R^2} = R^2 \left(1 - \frac{1}{1+\mu}\frac{f_S^2}{h_R^2}\right) \tag{52}$$

Finally, the squared position vector magnitude can be determined using (49) and (52) as

$$v_r^2 = x_{3D}^2 + y_{3D}^2 + z_{3D}^2 = R^2 \left(1 - \frac{\mu}{(1+\mu)^2}\frac{f_S^2}{h_R^2}\right) = \frac{R^2}{1+\mu}\left(2+\mu - \frac{1}{h_R^2}\right) \tag{53}$$

where (17) was used for the rightmost expression derivation.

When we return to Eq. (45), we see that the first determined non-trivial and non-constant solution of the angular part of the Laplace equation is

$$F_{A1}(W) = c_{A1}\frac{f_S}{h_R} \tag{54}$$

The $f_S/h_R$ function is here multiplied by the arbitrary constant $c_{A1}$, as such multiplication evidently also leads to a solution of (37). The index 1 denotes that this is the solution valid for the separation constant $K_d$ equal to 1, while the index $A$ denotes that this is the first-kind solution of the expected two different kinds of solution (as the equation to be solved is the second-order differential equation).

Note that for $\mu = 0$, $f_S/h_R$ results in $\sin(\nu)$. In this limit, the solution function is

$$F_0(W)|_{\mu=0} \sim \sin \nu \tag{55}$$

This function is basically the solution of the angular part of the Laplace equation in the spherical coordinates, particularly the first-degree Legendre polynomial (considering that we do not have polar coordinates—the origin of $\nu$ coordinate is here at the equator). It does worth to note that powers of $\sin(\nu)$ form (within the Legendre polynomials) the angular part of the solution of the Laplace equation in the spherical coordinates.

## 7 Simple solution of the second kind

It has to be noted that (54) is only a particular solution of Eq. (38). Nevertheless, the general solution for $K_d = 1$ can be obtained as well. The general solution (for $K_d = 1$) is given by (see [17])

$$F(W) = c_{A1} F_0(W) + c_{B1} F_0(W) \int \frac{1}{[F_0(W)]^2} \exp[-\int f(W)dW]dW = F_{A1}(W) + F_{B1}(W) \tag{56}$$

The first part, i.e., by $c_{A1}$, is a solution of its own, as it is the already found solution (54). The second part, i.e., the function $F_{B1}(W)$, is thus a new, separate, second-kind solution:

$$F_{B1}(W) = c_{B1} F_0(W) \int \frac{1}{[F_0(W)]^2} \exp[-\int f(W)dW]dW \tag{57}$$

To find it in an analytic form, we have to determine the integrals (inner and outer) in (57). The integration and then the derivation of the second-kind solution for $K_d = 1$ is done in Supplement B, Sect. B4 (the relations (14), (15), (17), and also (A75), (A1), (A2), (A3), (A4), (A5), (A7) from Appendix A, as well as Eq. (82) from [4] are employed in the supplement). The second possible real solution of the angular part of the separated Laplace equation for the given model ($K_d = 1$) is then derived in the form

$$F_{B1}(W) = c_{B1} \frac{f_S}{h_R} \left[ \ln\left( \frac{f_S}{(1+\mu)f_C} + \sqrt{1 + \frac{f_S^2}{(1+\mu)^2 f_C^2}} \right) - \sqrt{1 + \frac{(1+\mu)^2 f_C^2}{f_S^2}} \right] \tag{58}$$

or, equivalently, with inverse hyperbolic sine instead of logarithm,

$$F_{B1}(W) = c_{B1} \frac{f_S}{h_R} \left[ \sinh^{-1}\left( \frac{f_S}{(1+\mu)f_C} \right) - \sqrt{1 + \frac{(1+\mu)^2 f_C^2}{f_S^2}} \right] \tag{59}$$

(taking into account the well-known relation between the inverse hyperbolic sine and the logarithm functions).

A correctness of this more complex solution (denoted "the second-kind solution" in what follows) of the full Laplace Eq. (24), in which the found radial part solution (33) with $K_d = 1$ is used, is successfully tested in Supplement B Sect. B5 ((17), (10), (16), and also (A91), (A93), (A18), (A98), (A78), (A92), (A38) listed in Appendix A are employed there). It is confirmed that (58) (as well as (59)) is one of the solutions.

## 8 The found solutions rewritten to the form suitable for further generalization

The found second-kind solution for $K_d = 1$, i.e., (58), contains relations with $f_S/f_C$. Nevertheless, it can be written in terms of $f_S/h_R$ as well, as (see Eq. (A24) in Appendix A)

$$\frac{f_S^2}{f_C^2} = \frac{\frac{f_S^2}{h_R^2}}{1 - \frac{1}{1+\mu} \frac{f_S^2}{h_R^2}} \tag{60}$$

It is of advantage to define a new function

$$s \equiv \frac{f_S}{h_R} \tag{61}$$

The power series expressions of $s$ are recorded in (46) and (47). As $h_R^2$ is equal to 1 on the equator and to $1/(1+\mu)$ on the polar axis (see Eq. (91) in [4] and employ the fact that $h_R$ is constant on straight lines going through the origin), and as $f_S$ is 0 on the equator and 1 on the polar axis (see [10], Eqs. (114) and (115)), the ratio $f_S/h_R$ is 0 on the equator, while it is equal to $\sqrt{1+\mu}$ on the polar axis, i.e.,

$$s|_{\nu=0} = \left.\frac{f_S}{h_R}\right|_{\nu=0} = 0 \quad \text{and} \quad s|_{\nu=\pi/2} = \left.\frac{f_S}{h_R}\right|_{\nu=\pi/2} = \sqrt{1+\mu} \tag{62}$$

Further (see Eq. (91) in [4] and Eq. (123) in [10]), on the reference spheroid, $s$ can be expressed in a closed form as

$$s|_{R=R_0} = \frac{f_S}{h_R}\bigg|_{R=R_0} = \frac{f_{S0}}{h_{R0}} = \frac{\sqrt{\frac{1+\mu}{1+\mu \sin^2 \nu}} \sin^2 \nu}{\frac{1}{\sqrt{1+\mu \sin^2 \nu}}} = \sqrt{1+\mu} \sin \nu \tag{63}$$

With a help of (61), the second-kind solution (58) in terms of $s$ is (see the derivation in Supplement C, part C1)

$$F_{B1}(s) = c_{B1}\left[s\frac{1}{2}\ln\left(\frac{\left[s+\sqrt{(1+\mu)^2 - \mu s^2}\right]^2}{(1+\mu)\left[(1+\mu) - s^2\right]}\right) - \sqrt{(1+\mu)^2 - \mu s^2}\right] \tag{64}$$

or equivalently

$$F_{B1}(s) = c_{B1}\left[s\frac{1}{2}\ln\left(\frac{\sqrt{(1+\mu)^2 - \mu s^2} + s}{\sqrt{(1+\mu)^2 - \mu s^2} - s}\right) - \sqrt{(1+\mu)^2 - \mu s^2}\right] \tag{65}$$

Note that the expression in the square brackets simplifies, for $\mu = 0$ (i.e., for the spherical case), to

$$\left[s\frac{1}{2}\ln\left(\frac{1+s}{1-s}\right) - 1\right] \tag{66}$$

which is the first degree Legendre function of the second kind.

Similarly, the first determined solution of the angular part of the Laplace equation, i.e., the first-kind solution, is written in terms of $s$ as

$$F_{A1}(s) = c_{A1} s \tag{67}$$

which is (for $\mu = 0$) the Legendre function of the first kind (or Legendre polynomial) of the first degree (see Eq. (55)).

It has to be noted that the range of the variable $s$ is, in the SOS case, larger than in the spherical case (where the range is $<-1, 1>$). According to (62), the upper limit of $s$ is $\sqrt{1+\mu}$ in the SOS case.

## 9 The angular part of the Laplace equation rewritten to the form suitable for further generalization

We now know that solutions exist resembling Legendre functions of the first degree and both of the first kind and of the second kind. Therefore, it does worth to test if the angular part of the Laplace equation in the SOS coordinates can be written in the form similar to the Legendre equation.

Further step is thus to rewrite the angular part of the Laplace equation in the SOS coordinates (37) to the form containing only the variable $s$. The derivation is reported in Supplement C, part C2. With the help of the relations (A19), (A20) and (A31) from Appendix A, we arrive at the equation

$$\begin{aligned}&\left[(1+\mu) - s^2\right]\left[(1+\mu)^2 - \mu s^2\right]\frac{d^2 F(s)}{ds^2}\\ &+ s\left[-(3\mu + 2)(1+\mu) + 2\mu(1+\mu)K_d + \mu(3 - 2K_d)s^2\right]\frac{dF(s)}{ds}\\ &+ K_d\left[(K_d - 2)\mu s^2 + (1+\mu)K_d + (1+\mu)^2\right]F(s) = 0\end{aligned} \tag{68}$$

termed "generalized Legendre equation" in what follows. Indeed, for a special case when $\mu = 0$ (i.e., for the spherical case) the equation reduces to the well-known Legendre equation

$$(1 - s^2)\frac{d^2 F(s)}{ds^2} - 2s\frac{dF(s)}{ds} + K_d(K_d + 1)F(s) = 0 \tag{69}$$

## 10 Solutions for the special case $K_d = 0$

We already know the solution of (68) in the special case $K_d = 1$ (see Eq. (67) and (65)). We can relatively easily find the solutions in the case when the separation constant $K_d = 0$. (68) then simplifies to

$$\left[(1+\mu) - s^2\right]\left[(1+\mu)^2 - \mu s^2\right]\frac{d^2 F(s)}{ds^2} + s\left[-(3\mu + 2)(1+\mu) + 3\mu s^2\right]\frac{dF(s)}{ds} = 0 \tag{70}$$

It is clear that one solution (the first-kind solution) is when $F(s)$ equals to a constant, here denoted $c_{A0}$:

$$F_{A0}(s) = c_{A0} \tag{71}$$

The second-kind solution is more complicated. Nevertheless, if we use—as a test function—the part of the already found second-kind first-degree solution (64), particularly the logarithm-based function,

$$\frac{1}{2}\ln\left(\frac{\left[s+\sqrt{(1+\mu)^2-\mu s^2}\right]^2}{(1+\mu)\left[(1+\mu)-s^2\right]}\right) \tag{72}$$

we get its first and second derivatives as follows. The first derivative is

$$\frac{d}{ds}\left[\sinh^{-1}\left(\frac{1}{\sqrt{1+\mu}}\frac{s}{\sqrt{(1+\mu)-s^2}}\right)\right] = \frac{d}{ds}\left[\frac{1}{2}\ln\left(\frac{\left[s+\sqrt{(1+\mu)^2-\mu s^2}\right]^2}{(1+\mu)\left[(1+\mu)-s^2\right]}\right)\right]$$
$$= \frac{(1+\mu)}{\left((1+\mu)-s^2\right)}\frac{1}{\sqrt{(1+\mu)^2-\mu s^2}} \tag{73}$$

, and the second derivative is

$$\frac{d}{ds}\left\{\frac{d}{ds}\left[\frac{1}{2}\ln\left(\frac{\left[s+\sqrt{(1+\mu)^2-\mu s^2}\right]^2}{(1+\mu)\left[(1+\mu)-s^2\right]}\right)\right]\right\} = \frac{d}{ds}\left\{\frac{(1+\mu)}{\left((1+\mu)-s^2\right)}\frac{1}{\sqrt{(1+\mu)^2-\mu s^2}}\right\}$$
$$= \frac{(1+\mu)s}{\left((1+\mu)-s^2\right)^2}\frac{\left[(3\mu+2)(1+\mu)-3\mu s^2\right]}{\left[(1+\mu)^2-\mu s^2\right]\sqrt{(1+\mu)^2-\mu s^2}} \tag{74}$$

The derivatives inserted to (70) result in the equation

$$\left[(1+\mu)-s^2\right]\left[(1+\mu)^2-\mu s^2\right]\frac{(1+\mu)s}{\left((1+\mu)-s^2\right)^2}\frac{\left[(3\mu+2)(1+\mu)-3\mu s^2\right]}{\left[(1+\mu)^2-\mu s^2\right]\sqrt{(1+\mu)^2-\mu s^2}}$$
$$+s\left[-(3\mu+2)(1+\mu)+3\mu s^2\right]\frac{(1+\mu)}{\left((1+\mu)-s^2\right)}\frac{1}{\sqrt{(1+\mu)^2-\mu s^2}} = 0 \tag{75}$$

After cancellation of a number of terms in the equation, what remains is

$$\left[(3\mu+2)(1+\mu)-3\mu s^2\right]+\left[-(3\mu+2)(1+\mu)+3\mu s^2\right] = 0 \tag{76}$$

which is obviously fulfilled for any $s$. Then, the given function (72) multiplied by a constant, i.e.,

$$F_{B0}(s) = c_{B0}\frac{1}{2}\ln\left(\frac{\left[s+\sqrt{(1+\mu)^2-\mu s^2}\right]^2}{(1+\mu)\left[(1+\mu)-s^2\right]}\right) \tag{77}$$

is the searched solution of the second kind for $K_d = 0$.

## 11 Polynomials as a solution of the angular part of the Laplace equation in the SOS coordinates

It was shown in the previous text that $s = f_S/h_R$ function is a basis for the found simple solutions of the angular part of the Laplace equation in the SOS coordinates for $K_d = 0$ and $K_d = 1$. Moreover, $f_S/h_R$ function becomes $\sin(v)$ in the limit $\mu = 0$. At the same time, it is known that powers of $\sin(v)$ form (within the Legendre polynomials) the solution of the Laplace equation in the spherical coordinates.

Therefore, a possibility is further tested, if the angular part of the Laplace equation in the SOS coordinates for various $K_d$ could be written generally in the form of Legendre-like polynomials, which employ $f_S/h_R$ (i.e., the generalized sine divided by the metric scale factor of $R$ coordinate) instead of the standard sine as the argument, i.e.,

$$F_{A\beta_2}(s) = \sum_{j=\beta_1}^{\beta_2} c_j \left(\frac{f_S}{h_R}\right)^j = \sum_{j=\beta_1}^{\beta_2} c_j s^j \tag{78}$$

Here, $c_j$ are unknown constants, and $\beta_1$, $\beta_2$, $\beta_2 \geq \beta_1$, are some integer numbers determining the number and power of the solution terms. This model, hopefully, will lead to the first-kind solutions when proper $c_j$ constants are found.

The angular part of the Laplace Eq. (68) can be—with this model—written in the form

$$[(1+\mu)-s^2][(1+\mu)^2-\mu s^2]\frac{d^2 F_{A\beta_2}(s)}{ds^2}$$
$$+s[-(3\mu+2)(1+\mu)+2\mu(1+\mu)K_d+\mu(3-2K_d)s^2]\frac{dF_{A\beta_2}(s)}{ds} \quad (79)$$
$$+K_d[(K_d-2)\mu s^2+(1+\mu)K_d+(1+\mu)^2]F_{A\beta_2}(s)=0$$

The first and second derivatives of $F_{A\beta_2}(s)$ (which are easily acquired for a polynomial function), as well as the polynomial $F_{A\beta_2}(s)$ itself, are inserted to (79). The equation is subsequently simplified. This is done in Supplement D, part D1. Then, the following equation is obtained:

$$\sum_{j=\beta_1}^{\beta_2} c_j \{(1+\mu)^3 j(j-1)+(1+\mu)[-(1+2\mu)j^2+\{2\mu K_d-(1+\mu)\}j+K_d[K_d+(1+\mu)]]s^2 \quad (80)$$
$$+\mu[j^2+2(1-K_d)j+K_d(K_d-2)]s^4\} s^{j-2}=0$$

This is the angular part of the Laplace equation in the SOS coordinates when (78) polynomial is used as a solution model.

It can be seen that the term with the pre-factor $j(j-1)(1+\mu)^3$ in (80) will be always the term with the lowest power when $j$ is equal to $\beta_1$, and—moreover—it cannot be compensated by any other term in the sum as no other has that low power. Therefore, it has to be zero for $j$ equal to $\beta_1$. Similarly, this term cannot be compensated by any other term (as no other has that power of $s$) also for $j$ equal to $\beta_1+1$. This restricts the bottom limit of the sum, $\beta_1$, to 0. In such case ($\beta_1=0$), both abovementioned terms are zero thanks to the pre-factor $j(j-1)$. Thus, we select $\beta_1=0$ in what follows.

Further, it can be seen that the term with the fourth power in $s$ will always give in the sum the term with the largest power in $s = f_S/h_R$ when $j$ is equal to $\beta_2$. Moreover, it cannot be compensated by any other term in the sum as no other has that large power. Therefore, the pre-factor has to be zero itself (in order the particular $c_{\beta_2}$ could be nonzero). It can be easily shown that in order to fulfill

$$\mu[\beta_2^2+2(1-K_d)\beta_2+K_d(K_d-2)]=0 \quad (81)$$

, the roots have to be $\beta_2=K_d$ or $\beta_2=K_d-2$. Moreover, it can be seen that the term with the fourth power in $s$ in (80) will always give in the sum the term with the second largest power in $s$ when $j$ is equal to $\beta_2-1$. It cannot be compensated by any other term as no other has that large power. It can be easily shown that in order to fulfill

$$\mu[(\beta_2-1)^2+2(1-K_d)(\beta_2-1)+K_d(K_d-2)]=0 \quad (82)$$

, the roots have to be $\beta_2=K_d\pm 1$, or $c_{\beta_2-1}$ has to be zero. As the $\beta_2$ roots for the abovementioned cases $j=\beta_2$ and $j=\beta_2-1$ are mutually incompatible, it is clear that $c_{\beta_2-1}$ has to be zero, and that the upper limit of the sum, $\beta_2$, is either equal to $K_d$ or to $K_d-2$. We conservatively select the larger one, i.e., $K_d$. Moreover, as the upper limit of the sum, $\beta_2$, has to be an integer number, then $K_d$ has to be an integer, too. Further, it has to be positive or zero (i.e., $K_d\geq 0$), according to our condition $\beta_2\geq \beta_1$. Thus, $K_d$ cannot be negative. This is a significant result, as it is different from the spherical case (i.e., when $\mu=0$, also discussed in Supplement D part D1) for which negative constants $K_d$ are also allowed. It means that (78) model polynomial cannot lead to a solution for any negative separation constant $K_d$ when $\mu>0$. Further search for solutions with negative exponents $K_d$ in the radial part (i.e., in fact for the exterior points up to the infinity) has still to continue in the future for the spheroidal (SOS) case.

For integer $K_d\geq 0$, we further employ the above derived limits in the sum of (80), and we then have the basic equation for the model (78), i.e., for the first-kind solutions, in the form

$$\sum_{j=0}^{K_d} c_j \{(1+\mu)^3 j(j-1)+(1+\mu)[-(1+2\mu)j^2+\{2\mu K_d-(1+\mu)\}j+K_d[K_d+(1+\mu)]]s^2 \quad (83)$$
$$+\mu[j^2+2(1-K_d)j+K_d(K_d-2)]s^4\} s^{j-2}=0$$

## 12 Solutions of the first kind for the special cases $K_d = 2, 3, 4, 5$ and $6$

We already know the solution in the form of polynomial for $K_d=0$ and 1 (see Eqs. (71) and (67)). In Supplement D, part D2, the polynomial solutions of (83) for $K_d=2, 3, 4, 5$ and 6 are derived, and they are reported here.

For $K_d=2$, the polynomial solution is

$$F_{A2}(s)=\frac{1}{2}[(\mu+3)s^2-(1+\mu)^2], \quad (84)$$

while it is

$$F_{A3}(s) = \frac{1}{2}\left[(3\mu + 5)s^3 - 3(1+\mu)^2 s\right] \tag{85}$$

for $K_d = 3$. The polynomial solution for $K_d = 4$ is

$$F_{A4}(s) \sim \left[(3\mu^2 + 30\mu + 35)s^4 - 2 \cdot 3(\mu + 5)(1+\mu)^2 s^2 + 3(1+\mu)^4\right], \tag{86}$$

and the solution for $K_d = 5$ is

$$F_{A5}(s) \sim \left[(15\mu^2 + 70\mu + 63)s^5 - 10(3\mu + 7)(1+\mu)^2 s^3 + 15(1+\mu)^4 s^1\right] \tag{87}$$

Finally, the polynomial solution for $K_d = 6$ is

$$\begin{aligned}F_{A6}(s) \sim &\left[(5\mu^3 + 105\mu^2 + 315\mu + 231)s^6 - 5(3\mu^2 + 42\mu + 63)(1+\mu)^2 s^4\right.\\ &\left.+ 3 \cdot 5(\mu + 7)(1+\mu)^4 s^2 - 5(1+\mu)^6\right]\end{aligned} \tag{88}$$

Notice that—for $\mu = 0$—the derived polynomials reduce to the Legendre polynomials of degree 2 to 6. For the found polynomial solutions with various $K_d$, normalization factors which would enable employing a proper recursion formula have still to be determined. Such formula is the topic of the next section.

## 13 Generalized Legendre polynomials and Bonnet-like recursion formula

Derivation of the higher-degree polynomial solutions of the angular part of the separated Laplace equation in the SOS coordinates (similarly as in the previous section) is increasingly more and more time demanding for the extent of calculations involved. Therefore, it would be of advantage to have a recursion formula which can derive relatively easily the higher-degree polynomial from a limited number of the lower-degree polynomials. For the spherical coordinates case, such recursion is known as Bonnet formula. Its derivation is based on a formal expansion of the following generating function in powers of $t$,

$$\left[1 - 2xt + t^2\right]^{-\frac{1}{2}} = \sum_{n=0}^{\infty} P_n(x) t^n \tag{89}$$

where $P_n(x)$ are the Legendre polynomials. The Bonnet recursion formula for the spherical case Legendre polynomials is then

$$P_{n+1}(x) = \frac{2n+1}{n+1} x P_n(x) - \frac{n}{n+1} P_{n-1}(x) \tag{90}$$

For SOS coordinates, derivation based on the already known seven Legendre-like polynomials (for $K_d = 0, 1, 2, 3, 4, 5$ and 6, see the previous sections) results in the similar formula

$$P_{n+1}^{SI}(s) = \frac{2n+1}{n+1} \frac{1}{1+\mu} s P_n^{SI}(s) - \frac{n}{n+1}\left(1 - \frac{\mu}{(1+\mu)^2} s^2\right) P_{n-1}^{SI}(s) \tag{91}$$

which reproduces the first seven previously derived SOS Legendre-like polynomials $\left(n = K_d, \ P_n^{SI}(s) \sim F_{An}(s)\right)$, except different normalization factors. These polynomials are summarized in Table 1. We will call them the generalized Legendre polynomials (or generalized Legendre functions of the first kind), and use for them the notation $P_n^{SI}(s)$ where the upper index $SI$ discriminates them from the standard Legendre polynomials $P_n(s)$ applicable in the spherical case. The "$I$" part of the upper index tells that these polynomials are valid only for the interior solution of the angular part of the separated Laplace equation.

For $\mu = 0$, it can be easily seen that the recursion formula (91) reduces to the well-known Bonnet formula (90) for the spherical case Legendre polynomials.

In Supplement D, part D3, the recursion formula (91) is tested. It is used to determine higher-degree polynomials ($K_d = 2, 3, 4, 5$ and 6) from the first two ($K_d = 0, 1$). The results of the use of the Bonnet-like recursion formula (91) correspond to the ones determined in the previous section. The calculation by-products are the correct normalization coefficients of the polynomials, which are then included into the polynomials recorded in Table 1. In Table 1, therefore, the generalized Legendre polynomials $P_n^{SI}$ for the SOS coordinates with the proper normalization factors for the interior first-kind solution of the generalized Legendre equation up to 6th degree are listed.

It can be seen from Table 1 that for $\mu = 0$, the SOS case polynomial solutions are the same as for the spherical case (as $s = f_S/h_R$ is then equal to $\sin \nu = \sin \chi$). It can be also seen that the Legendre-like polynomials $P_n^{SI}$ are finite for any nonnegative $\mu$.

When the generalized Legendre polynomials $P_n^{SI}(s)$ are compared (see Table 1) with the Legendre polynomials $P_n(s)$ applicable for the spherical case ($\mu = 0$), we see that there are two main differences. First, the factors by the powers of $s$ are no more simple integers $k_0$, but they are more complex factors containing $\mu$, which are (without a mathematical rigor) of the form $(k_0 + k_1\mu + k_2\mu^2 + \ldots)(1+\mu)^l$. For $\mu = 0$, these factors result in the same integer number as for the spherical case Legendre polynomials, as expected. Further, for $\mu \neq 0$, a normalization factor exists differing from the spherical case one by $1/(1+\mu)^n$ for each generalized

**Table 1** Table of the first several lower-degree generalized Legendre polynomials $P_n^{Sl}(s)$ for SOS coordinates with the proper normalization factors for the first-kind solution of the generalized Legendre equation (i.e., the angular part of the separated Laplace equation in the SOS coordinates) up to the sixth degree

| $n$ | Polynomial $P_n^{Sl}(s)$ for the oblate spheroidal case | Polynomial $P_n(x)$ for the spherical case, $x = \sin\chi$ |
|---|---|---|
| 0 | $1$ | $1$ |
| 1 | $\frac{1}{1+\mu} \cdot s$ | $x$ |
| 2 | $\frac{1}{(1+\mu)^2} \cdot \frac{1}{2} \cdot \left[(\mu+3)s^2 - (1+\mu)^2\right]$ | $\frac{1}{2}\left(3x^2 - 1\right)$ |
| 3 | $\frac{1}{(1+\mu)^3} \cdot \frac{1}{2} \cdot \left[(3\mu+5)s^3 - 3(1+\mu)^2 s\right]$ | $\frac{1}{2}\left(5x^3 - 3x\right)$ |
| 4 | $\frac{1}{(1+\mu)^4} \cdot \frac{1}{2^3} \cdot \left[\left(3\mu^2 + 30\mu + 35\right)s^4 - 2 \cdot 3(\mu+5)(1+\mu)^2 s^2 + 3(1+\mu)^4\right]$ | $\frac{1}{8}\left(35x^4 - 2 \cdot 3 \cdot 5x^2 + 3\right)$ |
| 5 | $\frac{1}{(1+\mu)^5} \cdot \frac{1}{2^3} \cdot \left[\left(15\mu^2 + 70\mu + 63\right)s^5 - 10(3\mu+7)(1+\mu)^2 s^3 + 15(1+\mu)^4 s^1\right]$ | $\frac{1}{8}\left(63x^5 - 10 \cdot 7x^3 + 3 \cdot 5x\right)$ |
| 6 | $\frac{1}{(1+\mu)^6} \cdot \frac{1}{2^4} \cdot \left[\left(5\mu^3 + 105\mu^2 + 315\mu + 231\right)s^6 - 5\left(3\mu^2 + 42\mu + 63\right)(1+\mu)^2 s^4 + 3 \cdot 5(\mu+7)(1+\mu)^4 s^2 - 5(1+\mu)^6\right]$ | $\frac{1}{16}\left(231x^6 - 315x^4 + 105x^2 - 5\right)$ |

The substitution $s = f_S/h_R$ is used. All the polynomials of degree 0 to 6 were derived by a solution of the generalized Legendre equation. The polynomials of degree 2 to 6 were also checked by the Bonnet-like recursion formula (91). Comparison with the classical Legendre polynomials is provided in the last column. $\chi$ in the last column is latitude

Legendre polynomial. In the limit of the spherical case ($\mu \to 0$), the factor $1/(1+\mu)^n$ becomes equal to one for all the polynomials. For the spheroidal case ($\mu > 0$), nevertheless, it differs from 1 and, therefore, is essential when employing the recursion formula.

This above analysis indicates (although it is not a full rigorous proof for an arbitrary polynomial degree) that the generalized Legendre polynomials can be derived by the use of the recursion formula (91) also for the higher degrees than those listed in Table 1. It also indicates (although it is not proven rigorously) that the generalized Legendre polynomials determined in this way are solutions of the generalized Legendre equation (which is equivalent to the angular part of the Laplace equation in the SOS coordinates).

While the Legendre polynomials are orthogonal (with respect to the weighting function equal to 1), simple integration tests show that the generalized Legendre polynomials are not orthogonal with respect to the same weighting function. It is, therefore, more complicated to find proper coefficients for a solution given by particular boundary conditions. This is, nevertheless, out of the scope of this paper, and we leave finding of a proper weighting function on experienced mathematicians. (We only note that the known procedures used for Bessel functions weighting [18, 19] do not work as the $K_d$ separation constant is present in (68) equation also in the factor by the first derivative, not only by the function itself.)

## 14 The second-kind solutions of the second degree and of the third degree

Let us title the second solution of the generalized Legendre equation (or equivalently of the angular part of the Laplace equation in the SOS coordinates) the generalized Legendre function of the second kind. We will denote it $Q_n^S(s)$, where $n$ denotes degree. Two second-kind solutions were previously found: for $K_d = 0$ (see Eq. (77)), to which $Q_0^S(s)$ is proportional, i.e.,

$$Q_0^S(s) = \frac{1}{2} \ln\left( \frac{\left[s + \sqrt{(1+\mu)^2 - \mu s^2}\right]^2}{(1+\mu)\left[(1+\mu) - s^2\right]} \right) \tag{92}$$

and for $K_d = 1$ (see Eq. (64)), to which $Q_1^S(s)$ is proportional, i.e.,

$$Q_1^S(s) = \frac{1}{1+\mu}\left[ s \frac{1}{2} \ln\left( \frac{\left[s + \sqrt{(1+\mu)^2 - \mu s^2}\right]^2}{(1+\mu)\left[(1+\mu) - s^2\right]} \right) - \sqrt{(1+\mu)^2 - \mu s^2} \right] \tag{93}$$

The above relations for the generalized Legendre functions of the second kind are with the normalization factor set in such a way that the initial part of the relations equals to $P_0^{SI}(s)Q_0^{SI}(s)$ and $P_1^{SI}(s)Q_0^{SI}(s)$.

According to (62), the upper limit (i.e., on the pole) of $s$ is $\sqrt{1+\mu}$. When assessing the denominator value in the logarithm function, it is seen that the generalized Legendre functions of the second kind diverge on the axis of rotation, similarly as in the spherical case. Therefore, the second-kind solution is probably of less applicability in physics than the first-kind solution. Nevertheless, as there could be some special applications also for the second-kind solution in case of singularity, e.g., when solving the Laplace equation in Schwarzschild spacetime [20], it does worth to derive the higher-degree second-kind solutions of the generalized Legendre equation in the SOS case as well.

The second degree solution of the second kind was found by a tedious way of testing a model function. The derivation is done in Supplement E, part E1. The result is as follows:

$$Q_2^S(s) \sim \frac{(3+\mu)s^2 - (1+\mu)^2}{2} \frac{1}{2} \ln\left( \frac{\left[s + \sqrt{(1+\mu)^2 - \mu s^2}\right]^2}{(1+\mu)\left[(1+\mu) - s^2\right]} \right) - \frac{3}{2} s \sqrt{(1+\mu)^2 - \mu s^2} \tag{94}$$

It can be easily checked that—in the special case when $\mu = 0$—it would simplify to the classical second-kind Legendre function of the second degree.

For further test of a prospective recursion procedure, it does worth to have also the third degree solution. Extensive derivation and algebraic manipulations shown in detail in Supplement E, part E2, lead to the following formula for the solution:

$$Q_3^S(s) \sim \frac{1}{2}\left((5+3\mu)s^3 - 3(1+\mu)^2 s\right) \frac{1}{2} \ln\left( \frac{\left[s + \sqrt{(1+\mu)^2 - \mu s^2}\right]^2}{(1+\mu)\left[(1+\mu) - s^2\right]} \right) \\ - \frac{1}{2}\left( \frac{4\mu+15}{3} s^2 - \frac{4}{3}(1+\mu)^2 \right) \sqrt{(1+\mu)^2 - \mu s^2} \tag{95}$$

When $\mu = 0$, then the function simplifies to the classical second-kind Legendre function of the third degree, as can be easily found by the substitution $\mu = 0$:

$$\left. Q_3^S(s) \right|_{\mu=0} = Q_3(s) = \frac{1}{2}(5s^3 - 3s)\frac{1}{2}\ln\left(\frac{1+s}{1-s}\right) - \left(\frac{5}{2}s^2 - \frac{2}{3}\right) \tag{96}$$

## 15 The second-kind solutions

The already found second-kind solutions of the degree 0 to 3 are listed in Table 2. In Table 2, there are—for comparison—also listed the classical Legendre functions of the second kind for the spherical case up to the degree six. It has to be noted that standardly, these functions are reported only up to the degree 3. Therefore, $Q_4(x)$ and $Q_5(x)$ were obtained as the corresponding coefficient of expansion of the generating function

$$\left[1 - 2xt + t^2\right]^{-\frac{1}{2}} \sinh^{-1}\left(\frac{x-t}{\sqrt{1-x^2}}\right) = \sum_{n=0}^{\infty} Q_n(x) t^n \tag{97}$$

A generating function different from (97) which can help to determine the higher-degree Legendre functions of the second-kind $Q_n(x)$ is based on the generating function

$$\left[1 - 2xt + t^2\right]^{-\frac{1}{2}} \ln\left(\frac{-x + t + \left[1 - 2xt + t^2\right]^{\frac{1}{2}}}{1-x}\right) = \sum_{n=0}^{\infty} U_n(x) t^n \tag{98}$$

suggested in [21]. $U_n(x)$ functions are here the polynomials forming the second part (i.e., the one not by the logarithm) of the Legendre functions of the second kind, and thus

$$Q_n(x) = P_n(x) Q_0(x) - U_n(x) \tag{99}$$

The classical Legendre functions of the second-kind $Q_5(x)$ and $Q_6(x)$ in Table 2 were obtained using (99).

It is known that for the spherical case solution ($\mu = 0$), the Bonnet recursion formula (90) is valid also for the second-kind solutions. With its help, higher-degree second-kind Legendre functions can be relatively easily derived when $\mu = 0$. It indicates that the Bonnet-like formula (91) could similarly work also for the second-kind generalized Legendre functions, i.e., for the SOS coordinates case.

It can be deduced by comparing $P_n^{SI}(s)$ in Table 1 with Eqs. (92), (93), (94) and (95) that the following relation holds for the second-kind functions:

$$Q_n^{SI}(s) = P_n^{SI}(s) Q_0^{SI}(s) - T_n(s)\sqrt{(1+\mu)^2 - \mu s^2} \tag{100}$$

where $T_n(s)$ is a polynomial (especially, $T_0(s) = 0$, $T_1(s) = 1/(1+\mu)$). The first term in (100) has identical form as for the standard Legendre functions (see e.g., [22, 23]), and the second term contains special function $\sqrt{(1+\mu)^2 - \mu s^2}$ which equals to 1 for $\mu = 0$. In Supplement E part E3, a test is performed if the recursion formula (91) can be used (similarly as for the standard Legendre functions) also for the generalized Legendre function of the second kind, i.e.,

$$Q_{n+1}^{SI}(s) = \frac{2n+1}{n+1} \frac{1}{1+\mu} s Q_n^{SI}(s) - \frac{n}{n+1}\left(1 - \frac{\mu}{(1+\mu)^2} s^2\right) Q_{n-1}^{SI}(s) \tag{101}$$

and which normalization coefficients are to be used. The derivation confirmed that the recursion formula reproduces the found solutions for the second (Eq. (94)) and third (Eq. (95)) degree function. The functions with a proper normalization coefficients for the use in the recursion formula are reported in Table 2. Further, (101) is used in Supplement E, part E3, also for determination of the second-kind functions $Q_4^{SI}(s)$, $Q_5^{SI}(s)$ and $Q_6^{SI}(s)$ which are then also included into Table 2. It can be observed that $Q_n^{SI}(s) = Q_n(x)$ when $\mu = 0$ in all cases.

As the formula for $Q_n^{SI}(s)$ determination contains in the pre-factor by logarithm (i.e., by $Q_0^{SI}(s)$) the term $P_n^{SI}(s)$, which is the first-kind solution of the generalized Legendre equation, the range of the separation constant $K_d$ for which the second-kind solutions exist is the same as for the first-kind solutions. Therefore, neither the second-kind solution (similarly as the first-kind solutions) derived in the form shown above does exist for negative separation constants $K_d$.

## 16 The complete solution in interior points

Finally, the complete interior solution of the azimuthally symmetric case of the Laplace equation in orthogonal similar oblate spheroidal coordinates (20) with applied variables separation introduced by (23) can be written in the form

$$V(R, \nu) = V(R, W) = V\left(R, \frac{f_2}{h_R}\right) = V(R, s) = \sum_{n=0}^{\infty} a_n R^n P_n^{SI}(s) + \sum_{n=0}^{\infty} b_n R^n Q_n^{SI}(s) \tag{102}$$

where the radial part solution (33) is used, and $P_n^{SI}(s)$ and $Q_n^{SI}(s)$ are the generalized Legendre functions of the first and of the second kind, respectively. Several low-index functions can be seen in Tables 1 and 2. Note that the zero degree solution $a_0 R^0 P_0^{SI}(s)$ is equal to the constant $V_C$ (see Eq. (23)). $a_n$ and $b_n$ are arbitrary real coefficients, which can be determined for the particular boundary conditions. (102) thus represents the harmonic functions in the SOS coordinates.

**Table 2** Generalized Legendre functions of the second-kind $Q_n^{SI}$ applicable for the solution of the Laplace equation in the SOS coordinates up to the degree six with proper normalization factors

| $n$ | Function $Q_n^{SI}(s)$ for the oblate spheroidal case | Function $Q_n(x)$ for the spherical case |
|---|---|---|
| 0 | $Q_0^{SI}(s) = \frac{1}{2}\ln\left(\left[\frac{s+\sqrt{(1+\mu)^2-\mu s^2}}{(1+\mu)[(1+\mu)-s^2]}\right]^2\right)$ | $Q_0(x) = \frac{1}{2}\ln\left(\frac{1+x}{1-x}\right)$ |
| 1 | $P_1^{SI}(s)Q_0^{SI}(s) - \frac{1}{(1+\mu)}$ | $P_1(x)Q_0(x) - 1$ |
| 2 | $P_2^{SI}(s)Q_0^{SI}(s) - \frac{1}{(1+\mu)^2}\frac{1}{2}3s\sqrt{(1+\mu)^2 - \mu s^2}$ | $P_2(x)Q_0(x) - \frac{1}{2}3x$ |
| 3 | $P_3^{SI}(s)Q_0^{SI}(s) - \frac{1}{(1+\mu)^3}\frac{1}{2}\left[\left(\frac{4}{3}\mu+5\right)s^2 - \frac{4}{3}(1+\mu)^2\right]\sqrt{(1+\mu)^2 - \mu s^2}$ | $P_3(x)Q_0(x) - \frac{1}{2}\left(5x^2 - \frac{4}{3}\right)$ |
| 4 | $P_4^{SI}(s)Q_0^{SI}(s) - \frac{1}{(1+\mu)^4}\frac{1}{2^3}\left[\left(\frac{55}{3}\mu+35\right)s^3 - (1+\mu)^2\frac{55}{3}s\right]\sqrt{(1+\mu)^2 - \mu s^2}$ | $P_4(x)Q_0(x) - \frac{1}{8}\left(35x^3 - \frac{55}{3}x\right)$ |
| 5 | $P_5^{SI}(s)Q_0^{SI}(s) - \frac{1}{(1+\mu)^5}\frac{1}{2^3}\left[\left(\frac{64}{15}\mu^2 + 49\mu + 63\right)s^4 - \left(\frac{128}{15}\mu + 49\right)(1+\mu)^2 s^2 + \frac{64}{15}(1+\mu)^4\right]\sqrt{(1+\mu)^2 - \mu s^2}$ | $P_5(x)Q_0(x) - \frac{1}{8}\left(63x^4 - 49x^2 + \frac{64}{15}\right)$ |
| 6 | $P_6^{SI}(s)Q_0^{SI}(s) - \frac{1}{(1+\mu)^6}\frac{1}{2^4}\left[\left(\frac{231}{5}\mu^2 + 238\mu + 231\right)s^5 - \left(\frac{462}{5}\mu^2 + 238\mu + 238\right)(1+\mu)^2 s^3 + \frac{231}{5}(1+\mu)^4 s\right]\sqrt{(1+\mu)^2 - \mu s^2}$ | $P_6(x)Q_0(x) - \frac{1}{16}\left(231x^5 - 238x^3 + \frac{231}{5}x\right)$ |

The substitution $s = fs/h_R$ is used. Comparison with the Legendre functions of the second-kind $Q_n$ (the last column) is done as well. The polynomials $P_n^{SI}(s)$ and $P_n(x)$ appearing in the table can be found in Table 1

However, as shown in the previous text, this solution exists only for the nonnegative integer values of the separation constant $K_d = n$.

## 17 Conclusions

The interior solution of the azimuthally symmetric case of the Laplace equation in the orthogonal similar oblate spheroidal coordinates (20) was found (see Eq. (102)). During the derivation, the Laplace equation was separated by a special separation procedure to the radial (see Eqs. (31), (32)) and the so-called angular (see Eq. (36)) parts. While the solution (33) of the radial part is simple power of the equatorial radius of the oblate spheroidal coordinate surfaces, the separated angular part of the Laplace equation is significantly more complicated and leads to the generalized Legendre Eq. (68). Its solution then leads to the generalized Legendre functions of the first and of the second kind with argument $s = f_S/h_R$, i.e., the generalized sine divided by the metric scale factor of the coordinate $R$. Analytical expressions with infinite power series involving generalized binomial coefficients were written for this argument (see Eqs. (46) and (47)).

Several lower-degree generalized Legendre functions of both kinds were derived and are listed in Tables 1 and 2. The recursion formula for the generalized Legendre polynomials (see Eq. (91)) was found. The general solution of the rotationally symmetric Laplace equation for the interior space is reported (see Eq. (102)).

The presented solution is a step forward in the establishment of the SOS coordinates as a useful tool in the field physics. Further, the determination of the harmonic functions could facilitate finding solutions of more complex differential equations in the SOS coordinates, for example the Poisson equation or the Helmholtz equation.

Within the Laplace solution determination process, relevant identities valid for the SOS coordinate system in general and for the generalized sine and cosine functions in particular were derived as well (see Appendix A).

The found solution exists only for the nonnegative integer values of the exponent of the equatorial radius in the solution. Therefore, the solution has physical meaning only for the interior space (i.e., it is restricted to a certain annulus) and cannot be used for a region extended up to the infinity, as the solution would diverge. In order to find solutions in the exterior points, another approach has to be used. The exterior points solution is in fact already partially known to the author. However, due to the extent of the present text, it will be published elsewhere together with other considerations on the Laplace equation solution in the SOS coordinates.

**Supplementary Information** The online version contains supplementary material available at https://doi.org/10.1140/epjp/s13360-024-05181-4.

**Acknowledgements** The author acknowledges support from the long-term conceptual development project RVO 61389005 of the Nuclear Physics Institute of the Czech Academy of Sciences and from the OP JAK project of MEYS Nr. CZ.02.01.01/00/22_008/0004591. The author warmly thanks Věra Strunzová for her perpetual support.

**Data availability** This article deals with the derivation of theoretical relations within the similar oblate spheroidal coordinate system. There are thus no data sets to be disclosed.

**Declarations**

**Conflicts of interest** Authors declares no conflict of interest.

## Appendix A

### Derivation of relations required for the solution of the Laplace equation in SOS coordinates

The derivations in the following sections and subsections intend to find identities between some functions relevant in the SOS coordinate system, and also to express the derivatives of functions relevant in the SOS coordinate system with the help of the already known functions.

The structuring of the sections and subsections in the appendix is as follows:
A1. Pólya-Szegő identities.
A2. Basic operations with the expressions containing Pólya-Szegő-type power series.

A2.1. Jacobian divided by the square of the $h_\nu$ scale factor for small-$\nu$ region.

A2.2. Relations between metric scale factor $h_R$ and the generalized sine and cosine.

A2.3. Relations between $W$, $h_R$, $f_S$ and $f_C$.

A2.4. Relations involving $h_\nu$ metric scale factor.

A3. Derivatives of $W$ and $F(W)$ with respect to $\nu$ and $R$.

A3.1. Derivatives of $W$ with respect to $\nu$ and $R$.

A3.2. Derivatives of $F(W)$ with respect to $W$ and $\nu$.

A4. Derivatives of the Pólya-Szegő-type power series.

A4.1. Derivative of $S_C$ type of Pólya-Szegő power series.

A4.2. Derivative of $h_R^2$.

A4.3. Derivative of $S_A$ type of Pólya-Szegő power series.

A4.4. Summary for the derivative of $S_C$ and $S_A$ type of Pólya-Szegő power series.

A4.5. Derivatives of power series contained in metric scale factors.

A4.6. Derivatives of the generalized cosine and sine.

A4.7. Derivatives of functions containing Jacobian.

A5. Integral identities involving the Pólya-Szegő-type power series.

## A1. Pólya-Szegő identities

The essential identities, which the analytic relations of the orthogonal SOS coordinates are based on, are the Pólya and Szegő relations [11, 12, 15] or see [10], Eqs. (42), (43), (44)), i.e.,

$$\frac{p^{a+1}}{(1-b)p+b} = \sum_{k=0}^{\infty} \binom{a+bk}{k}(W^\varepsilon)^k \tag{A1}$$

$$p^a = \sum_{k=0}^{\infty} \frac{a}{a+bk}\binom{a+bk}{k}(W^\varepsilon)^k \tag{A2}$$

for which the following equation for $p$ is fulfilled:

$$-W^\varepsilon p^b + p = 1 \tag{A3}$$

The power series in (A1) and (A2) employ generalized binomial coefficients (see e.g., [10], Eq. (41)). For the SOS coordinates, particularly,

$$\varepsilon = 2, b = -\mu \tag{A4}$$

in the small-$\nu$ region, whereas

$$\varepsilon = -\frac{2}{1+\mu}, b = \frac{\mu}{1+\mu} \tag{A5}$$

in the large-$\nu$ region.

## A2. Basic operations with the expressions containing Pólya-Szegő-type power series

A2.1. Jacobian divided by the square of the **$h_\nu$** metric scale factor

For the use in the Laplace equation solution, we derive here the Jacobian divided by the square of the $h_\nu$ scale factor, first in the small-$\nu$ region. This quantity is needed in the case of Laplacian calculation in the SOS coordinates. We start with a division of (11) by the squared (8):

$$\begin{aligned}\frac{\Im}{h_\nu^2} &= \frac{\frac{R^2}{\sqrt{1+\mu}}\frac{\partial W}{\partial \nu}\sum_{k=0}^{\infty}\binom{-\frac{\mu+3}{2}-\mu k}{k}(W^2)^k}{\frac{R^2}{1+\mu}\left(\frac{\partial W}{\partial \nu}\right)^2\sum_{k=0}^{\infty}\binom{-(\mu+2)-\mu k}{k}(W^2)^k} \\ &= \sqrt{1+\mu}\,\frac{\sum_{k=0}^{\infty}\binom{-\frac{\mu+3}{2}-\mu k}{k}(W^2)^k}{\frac{\partial W}{\partial \nu}\sum_{k=0}^{\infty}\binom{-(\mu+2)-\mu k}{k}(W^2)^k}\end{aligned} \tag{A6}$$

For the determination of the ratio of the power series in (A6), the combination of the Hagen–Rothe identity with the Cauchy product (see Eq. (50) in [10])

$$\frac{\sum_{k=0}^{\infty}\binom{a+c+bk}{k}(W^{\varepsilon})^k}{\left[\sum_{k=0}^{\infty}\binom{c+bk}{k}(W^{\varepsilon})^k\right]} = \left[\sum_{k=0}^{\infty}\frac{a}{a+bk}\binom{a+bk}{k}(W^{\varepsilon})^k\right] \tag{A7}$$

is to be used. Here, $\varepsilon = 2$, $b = -\mu$, $c = -(\mu + 2)$ and $a = (\mu + 1)/2$, thus $a + c = -(\mu + 3)/2$. Then

$$\frac{\Im}{h_{\nu}^2} = \frac{\sqrt{1+\mu}}{\frac{\partial W}{\partial \nu}} \sum_{k=0}^{\infty} \frac{\frac{\mu+1}{2}}{\frac{\mu+1}{2} - \mu k} \binom{\frac{\mu+1}{2} - \mu k}{k}(W^2)^k \tag{A8}$$

$\Im/h_{\nu}^2$ ratio derivation for the large-$\nu$ region is carried out in a similar way by using (A7) with $\varepsilon = -2/(1+\mu)$, $b = \mu/(1+\mu)$, $c = -(2+\mu)/(1+\mu)$ and $a = 1/2$, thus $a + c = -(\mu + 3)/[2(1+\mu)]$. We arrive at

$$\frac{\Im}{h_{\nu}^2} = \frac{W\sqrt{1+\mu}}{\frac{\partial W}{\partial \nu}} \sum_{k=0}^{\infty} \frac{\frac{1}{2}}{\frac{1}{2} + \frac{\mu}{1+\mu}k} \binom{\frac{1}{2} + \frac{\mu}{1+\mu}k}{k}\left(W^{-\frac{2}{1+\mu}}\right)^k \tag{A9}$$

A2.2. Relations between metric scale factor $h_R$ and the generalized sine and cosine

Several relations between the metric scale factors and the generalized sine and cosine (see Eqs. (7) and (14), (15)) are needed for the Laplace equation solution. An important relation is $(1 - h_R^2)/\mu$, which can be—for the small-$\nu$ region—written as

$$\frac{1 - h_R^2}{\mu} = \frac{1 - \sum_{k=0}^{\infty}\binom{-\mu k}{k}(W^2)^k}{\mu}$$

$$= \frac{1 - 1 - \sum_{k=1}^{\infty}\binom{-\mu k}{k}(W^2)^k}{\mu} = \frac{-\sum_{M=0}^{\infty}\binom{-\mu(M+1)}{M+1}(W^2)^{M+1}}{\mu} \tag{A10}$$

where we used the substitution $M = k-1$. When we further use (subsequently) two well-known binomial identities

$$\binom{\alpha}{k} = \binom{\alpha}{k-1}\frac{\alpha-k+1}{k} \text{ and } \binom{\beta+M}{M}\frac{1}{\beta+M} = \binom{\beta+M-1}{M}\frac{1}{\beta} \tag{A11}$$

(see, e.g., [10], Eqs. (53) and (54)), we arrive at

$$\frac{1 - h_R^2}{\mu} = \frac{-W^2 \sum_{M=0}^{\infty}\binom{-\mu(M+1)}{M}\frac{-\mu(M+1)-(M+1)+1}{M+1}(W^2)^M}{\mu}$$

$$= W^2 \sum_{M=0}^{\infty}\binom{-\mu(M+1)}{M}\frac{-\mu(M+1)-M}{-\mu(M+1)}(W^2)^M$$

$$= W^2 \sum_{M=0}^{\infty}\binom{-\mu(M+1)-1}{M}\frac{-\mu(M+1)-M}{-\mu(M+1)-M}(W^2)^M = W^2 \sum_{M=0}^{\infty}\binom{-(1+\mu)-\mu M}{M}(W^2)^M \tag{A12}$$

If we use the notation coming from (14), which introduced the generalized sine $f_S$, in the last part of (A12), we obtain the relation

$$\frac{1 - h_R^2}{\mu} = \frac{f_S^2}{1+\mu} \tag{A13}$$

Similarly, for the large-$\nu$ region, we can write

$$\frac{1 - h_R^2}{\mu} = \frac{1 - \frac{1}{1+\mu}\sum_{k=0}^{\infty}\binom{\frac{\mu}{1+\mu}k}{k}\left(W^{-\frac{2}{1+\mu}}\right)^k}{\mu}$$

$$= \frac{1 - \frac{1}{1+\mu} - \frac{1}{1+\mu}\sum_{k=1}^{\infty}\binom{\frac{\mu}{1+\mu}k}{k}\left(W^{-\frac{2}{1+\mu}}\right)^k}{\mu}$$

$$= \frac{\frac{\mu}{1+\mu} - \frac{1}{1+\mu}\sum_{M=0}^{\infty}\binom{\frac{\mu}{1+\mu}(M+1)}{M+1}\left(W^{-\frac{2}{1+\mu}}\right)^{M+1}}{\mu} \tag{A14}$$

where we used the substitution $M = k-1$. When we further use (subsequently) the binomial identities listed in (A11), we get

$$\begin{aligned}\frac{1-h_R^2}{\mu} &= \frac{\frac{\mu}{1+\mu} - \frac{1}{1+\mu}\sum_{M=0}^{\infty}\binom{\frac{\mu}{1+\mu}(M+1)}{M}\frac{\frac{\mu}{1+\mu}(M+1)-(M+1)+1}{M+1}\left(W^{-\frac{2}{1+\mu}}\right)^{M+1}}{\mu}\\ &= \frac{1}{1+\mu} - \frac{1}{(1+\mu)^2}\sum_{M=0}^{\infty}\binom{\frac{\mu}{1+\mu}(M+1)}{M}\frac{\frac{\mu}{1+\mu}(M+1)-M}{\frac{\mu}{1+\mu}(M+1)}\left(W^{-\frac{2}{1+\mu}}\right)^{M+1}\\ &= \frac{1}{1+\mu}\left[1 - \frac{1}{1+\mu}W^{-\frac{2}{1+\mu}}\sum_{M=0}^{\infty}\binom{\frac{\mu}{1+\mu}(M+1)-1}{M}\frac{\frac{\mu}{1+\mu}(M+1)-M}{\frac{\mu}{1+\mu}(M+1)-M}\left(W^{-\frac{2}{1+\mu}}\right)^{M}\right]\\ &= \frac{1}{1+\mu}\left[1 - \frac{1}{1+\mu}W^{-\frac{2}{1+\mu}}\sum_{M=0}^{\infty}\binom{\frac{\mu}{1+\mu}(M+1)-1}{M}\left(W^{-\frac{2}{1+\mu}}\right)^{M}\right]\end{aligned} \quad (A15)$$

Using the definitions of the generalized cosine (14) as well as (16), we arrive at

$$\frac{1-h_R^2}{\mu} = \frac{1}{1+\mu}[1-f_C^2] = \frac{f_S^2}{1+\mu} \quad (A16)$$

(A13) and (A16) are formally the same. Therefore, the same relation can be used in both the small-$\nu$ region and the large-$\nu$ region.

For completeness, we report also the expression (derived with the help of Eq. (91) of [4]) valid at the reference surface (i.e., for $R = R_0$):

$$\frac{1-h_{R0}^2}{\mu} = \frac{\sin^2\nu}{1+\mu\sin^2\nu} \quad (A17)$$

By using (16) and (17) relations, the following relations valid in both regions can be easily derived between $h_R$ and the generalized sine and cosine:

$$h_R^2 - f_C^2 = h_R^2 - \frac{h_R^2}{\mu} - h_R^2 + \frac{1}{\mu} = \frac{1-h_R^2}{\mu} = \frac{f_S^2}{1+\mu} \quad (A18)$$

Further useful relations between the metric scale factor $h_R$ and and $f_S/h_R$ or $f_C/h_R$ are as follows:

$$\begin{aligned}\frac{f_S^2}{h_R^2} &= \frac{1+\mu}{\mu}\frac{(1-h_R^2)}{h_R^2} = \frac{1+\mu}{\mu}\left(\frac{1}{h_R^2}-1\right)\\ &\Rightarrow \frac{1}{h_R^2} = \frac{\mu}{1+\mu}\frac{f_S^2}{h_R^2}+1 \Rightarrow h_R^2 = \frac{1}{\frac{\mu}{1+\mu}\left(\frac{f_S}{h_R}\right)^2+1}\end{aligned} \quad (A19)$$

Thanks to (A19), we also obtain the following relation between $h_R$ and $f_S/h_R$:

$$\begin{aligned}(2+\mu)h_R^2 - 1 &= (2+\mu)\frac{1}{\frac{\mu}{1+\mu}\left(\frac{f_S}{h_R}\right)^2+1} - 1\\ &= \frac{(2+\mu)-\frac{\mu}{1+\mu}\left(\frac{f_S}{h_R}\right)^2-1}{\frac{\mu}{1+\mu}\left(\frac{f_S}{h_R}\right)^2+1}\\ &= \frac{(1+\mu)-\frac{\mu}{1+\mu}\left(\frac{f_S}{h_R}\right)^2}{\frac{\mu}{1+\mu}\left(\frac{f_S}{h_R}\right)^2+1} = \frac{(1+\mu)-\frac{\mu}{1+\mu}\left(\frac{f_S}{h_R}\right)^2}{\left[\frac{1}{h_R^2}\right]}\end{aligned} \quad (A20)$$

Further,

$$\begin{aligned}\mu f_C^2 = (1+\mu)h_R^2 - 1 &= (1+\mu)\frac{1}{\frac{\mu}{1+\mu}\left(\frac{f_S}{h_R}\right)^2+1} - 1\\ &= \frac{(1+\mu)-\frac{\mu}{1+\mu}\left(\frac{f_S}{h_R}\right)^2-1}{\frac{\mu}{1+\mu}\left(\frac{f_S}{h_R}\right)^2+1} = \frac{\mu-\frac{\mu}{1+\mu}\left(\frac{f_S}{h_R}\right)^2}{\frac{\mu}{1+\mu}\left(\frac{f_S}{h_R}\right)^2+1} = \mu\frac{1-\frac{1}{1+\mu}\left(\frac{f_S}{h_R}\right)^2}{\frac{\mu}{1+\mu}\left(\frac{f_S}{h_R}\right)^2+1}\end{aligned} \quad (A21)$$

Another useful relation comes from (17):

$$f_C^2 f_S^2 = \frac{(1+\mu)h_R^2 - 1}{\mu} \frac{1+\mu}{\mu}(1 - h_R^2) = \frac{1+\mu}{\mu^2}\left[(2+\mu)h_R^2 - (1+\mu)h_R^4 - 1\right] \quad (A22)$$

A very important relation is the one between $f_S/f_C$ and $f_S/h_R$. The derivation employs (A19). We obtain from (17)

$$\frac{f_S^2}{f_C^2} = \frac{\frac{1+\mu}{\mu}(1-h_R^2)}{\frac{(1+\mu)h_R^2-1}{\mu}} = \frac{(1+\mu)\left(1 - \frac{1}{\frac{\mu}{1+\mu}\left(\frac{f_S}{h_R}\right)^2+1}\right)}{(1+\mu)\frac{1}{\frac{\mu}{1+\mu}\left(\frac{f_S}{h_R}\right)^2+1} - 1}$$

$$= \frac{(1+\mu)\left(\frac{\mu}{1+\mu}\left(\frac{f_S}{h_R}\right)^2 + 1\right) - (1+\mu)}{(1+\mu) - \left(\frac{\mu}{1+\mu}\left(\frac{f_S}{h_R}\right)^2 + 1\right)} = \frac{\mu\left(\frac{f_S}{h_R}\right)^2}{\mu - \frac{\mu}{1+\mu}\left(\frac{f_S}{h_R}\right)^2} \quad (A23)$$

Thus,

$$\frac{f_S^2}{f_C^2} = \frac{(1+\mu)\frac{f_S^2}{h_R^2}}{(1+\mu) - \frac{f_S^2}{h_R^2}} \quad (A24)$$

We can also derive Cartesian coordinates in 3D as function of $f_S/h_R$. With the help of Eq. (47) of [4], of the relations (A2), (A7) and (A19), as well as of (14) and (16) from the main text of the present article, we can write squared $x_{3D}$ in the small-$\nu$ region as

$$x_{3D}^2 = \cos^2 \lambda R^2 \left[\sum_{k=0}^{\infty} \binom{-\frac{1}{2} - \mu k}{k} \frac{-\frac{1}{2}}{-\frac{1}{2} - \mu k}(W^2)^k\right]^2$$

$$= \cos^2 \lambda R^2 \left[p^{-\frac{1}{2}}\right]^2 = \cos^2 \lambda R^2 \sum_{k=0}^{\infty} \binom{-1 - \mu k}{k} \frac{-1}{-1 - \mu k}(W^2)^k \quad (A25)$$

$$= \cos^2 \lambda R^2 \frac{\sum_{k=0}^{\infty}\binom{-1-\mu k}{k}(W^2)^k}{\sum_{k=0}^{\infty}\binom{-\mu k}{k}(W^2)^k} = \cos^2 \lambda R^2 \frac{f_C^2}{h_R^2} = \cos^2 \lambda R^2 \left(1 - \frac{1}{1+\mu}\frac{f_S^2}{h_R^2}\right)$$

Accordingly, squared $y_{3D}$ can be derived as

$$y_{3D}^2 = \sin^2 \lambda R^2 \frac{f_C^2}{h_R^2} = \sin^2 \lambda R^2 \left(1 - \frac{1}{1+\mu}\frac{f_S^2}{h_R^2}\right) \quad (A26)$$

### A2.3. Relations between $W$, $h_R$, $f_S$ and $f_C$

When we look at the form of the generalized sine and cosine (14), (15) and use (A2), (A7), we see that—in the small-$\nu$ region—we have

$$W^{-\frac{1}{\mu}}\left(\frac{f_S}{f_C}\right)^{\frac{\mu+1}{\mu}} = W^{-\frac{1}{\mu}}\left(\frac{W\sqrt{1+\mu}\sqrt{\sum_{k=0}^{\infty}\binom{-(1+\mu)-\mu k}{k}(W^2)^k}}{\sqrt{\sum_{k=0}^{\infty}\binom{-1-\mu k}{k}(W^2)^k}}\right)^{\frac{\mu+1}{\mu}}$$

$$= W\left(\sqrt{1+\mu}\sqrt{\sum_{k=0}^{\infty}\frac{-\mu}{-\mu - \mu k}\binom{-\mu - \mu k}{k}(W^2)^k}\right)^{\frac{\mu+1}{\mu}}$$

$$= W\left(\sqrt{1+\mu}\sqrt{p^{-\mu}}\right)^{\frac{\mu+1}{\mu}} = W\left(\sqrt{1+\mu}\right)^{\frac{\mu+1}{\mu}}\sqrt{p^{-(1+\mu)}}$$

$$= W\left(\sqrt{1+\mu}\right)^{\frac{\mu+1}{\mu}}\sqrt{\sum_{k=0}^{\infty}\frac{-(1+\mu)}{-(1+\mu) - \mu k}\binom{-(1+\mu)-\mu k}{k}(W^2)^k}$$

$$= W\left(\sqrt{1+\mu}\right)^{1+\frac{1}{\mu}} \sqrt{\frac{\sum_{k=0}^{\infty} \binom{-(1+\mu) - \mu k}{k}(W^2)^k}{\sum_{k=0}^{\infty} \binom{-\mu k}{k}(W^2)^k}} = \left(\sqrt{1+\mu}\right)^{\frac{1}{\mu}} \frac{f_S}{h_R} \tag{A27}$$

In the large-$\nu$ region, we obtain in a similar way

$$W^{-\frac{1}{\mu}} \left(\frac{f_S}{f_C}\right)^{\frac{\mu+1}{\mu}} = W^{-\frac{1}{\mu}} \left(\frac{\sqrt{\sum_{k=0}^{\infty} \binom{-1+\frac{\mu}{1+\mu}k}{k}\left(W^{-\frac{2}{1+\mu}}\right)^k}}{\frac{W^{-\frac{1}{1+\mu}}}{\sqrt{1+\mu}}\sqrt{\sum_{k=0}^{\infty} \binom{-\frac{1}{1+\mu}+\frac{\mu}{1+\mu}k}{k}\left(W^{-\frac{2}{1+\mu}}\right)^k}}\right)^{\frac{\mu+1}{\mu}}$$

$$= \left(\sqrt{1+\mu}\right)^{\frac{\mu+1}{\mu}} \left(\sqrt{\sum_{k=0}^{\infty} \frac{-\frac{\mu}{1+\mu}}{-\frac{\mu}{1+\mu}+\frac{\mu}{1+\mu}k}\binom{-\frac{\mu}{1+\mu}+\frac{\mu}{1+\mu}k}{k}\left(W^{-\frac{2}{1+\mu}}\right)^k}\right)^{\frac{\mu+1}{\mu}}$$

$$= \left(\sqrt{1+\mu}\right)^{\frac{\mu+1}{\mu}} \left(\sqrt{p^{-\frac{\mu}{1+\mu}}}\right)^{\frac{\mu+1}{\mu}} = \left(\sqrt{1+\mu}\right)^{\frac{\mu+1}{\mu}} \sqrt{p^{-1}}$$

$$= \left(\sqrt{1+\mu}\right)^{\frac{\mu+1}{\mu}} \sqrt{\sum_{k=0}^{\infty} \frac{-1}{-1+\frac{\mu}{1+\mu}k}\binom{-1+\frac{\mu}{1+\mu}k}{k}\left(W^{-\frac{2}{1+\mu}}\right)^k}$$

$$= \left(\sqrt{1+\mu}\right)^{1+\frac{1}{\mu}} \sqrt{\frac{\sum_{k=0}^{\infty} \binom{-1+\frac{\mu}{1+\mu}k}{k}\left(W^{-\frac{2}{1+\mu}}\right)^k}{\sum_{k=0}^{\infty} \binom{\frac{\mu}{1+\mu}k}{k}\left(W^{-\frac{2}{1+\mu}}\right)^k}} = \left(\sqrt{1+\mu}\right)^{\frac{1}{\mu}} \frac{f_S}{h_R} \tag{A28}$$

It does worth to note that this expression works also for $\mu = 0$, as $\left(\sqrt{1+\mu}\right)^{\frac{1}{\mu}}$ has the limit for $\mu \to 0$ which is square root of $e$. In this limit, the expression is thus $e \sin \nu$. Therefore, we can write an important relation between the generalized trigonometric functions, the metric scale factor $h_R$ and the parameter $W$, valid in both regions, as

$$W^{-\frac{1}{\mu}} \left(\frac{f_S}{f_C}\right)^{\frac{\mu+1}{\mu}} = \left(\sqrt{1+\mu}\right)^{\frac{1}{\mu}} \frac{f_S}{h_R} \tag{A29}$$

The combination of (A24) with (A29) leads then to a possibility to express $W$ in terms of $f_S/h_R$:

$$W^{-\frac{1}{\mu}} \left(\frac{(1+\mu)\frac{f_S^2}{h_R^2}}{(1+\mu) - \frac{f_S^2}{h_R^2}}\right)^{\frac{\mu+1}{2\mu}} = \left(\sqrt{1+\mu}\right)^{\frac{1}{\mu}} \frac{f_S}{h_R} \Rightarrow W^{-2} \left(\frac{(1+\mu)}{(1+\mu) - \frac{f_S^2}{h_R^2}}\right)^{\mu+1} \left(\frac{f_S}{h_R}\right)^2 = (1+\mu) \tag{A30}$$

Then (using the definition $s \equiv f_S/h_R$),

$$W^2 = \frac{(1+\mu)^\mu \left(\frac{f_S}{h_R}\right)^2}{\left((1+\mu) - \left(\frac{f_S}{h_R}\right)^2\right)^{1+\mu}} = \frac{(1+\mu)^\mu s^2}{\left((1+\mu) - s^2\right)^{1+\mu}} = \frac{\frac{1}{1+\mu}s^2}{\left(1 - \frac{1}{1+\mu}s^2\right)^{1+\mu}} \tag{A31}$$

It can be also easily found by substituting (17) to (A31) that we can express $W^2$ using only $h_R$,

$$W^2 = \frac{(1-h_R^2)(\mu h_R^2)^\mu}{\left[(1+\mu)h_R^2 - 1\right]^{\mu+1}} \tag{A32}$$

which, with a subsequent use of (17), leads to

$$W^2 = \frac{\frac{\mu}{1+\mu} f_S^2 (\mu h_R^2)^\mu}{[\mu f_C^2]^{\mu+1}} = \frac{(h_R^2)^\mu}{1+\mu} \frac{f_S^2}{[f_C^2]^{1+\mu}} \tag{A33}$$

The $W$ definition (5), and a subsequent use of the binomial theorem, results in the identity

$$\frac{(h_R^2)^\mu}{1+\mu} \frac{f_S^2}{[f_C^2]^{1+\mu}} = \left(\frac{R}{R_0}\right)^{2\mu} \frac{\sin^2 \nu}{[\cos^2 \nu]^{1+\mu}} = \left(\frac{R}{R_0}\right)^{2\mu} \sum_{k=0}^{\infty} \binom{-(1+\mu)}{k}(-1)^k (\sin^2 \nu)^{k+1} \tag{A34}$$

### A2.4. Relations involving $h_\nu$ metric scale factor

We will also need relations involving $h_\nu$ metric scale factor. For the further derivations of derivatives, it does worth to define the relations with power series contained in the $h_\nu$ metric scale factor (8) by

$$S_\nu \equiv \sum_{k=0}^{\infty} \binom{-(\mu+2)-\mu k}{k}(W^2)^k \text{ and } S_\nu \equiv \frac{1}{1+\mu}\left(W^{-2-\frac{2}{1+\mu}}\right)\sum_{k=0}^{\infty}\binom{-\frac{2+\mu}{1+\mu}+\frac{\mu}{1+\mu}k}{k}\left(W^{-\frac{2}{1+\mu}}\right)^k \tag{A35}$$

where the left side relation holds for the small-$\nu$ region, and the right side relation for the large-$\nu$ region. Such definition makes the following derivations easier to overview and having lower number of multiplication factors. It also does worth to notice that $S_\nu$ depends only on $W$, not separately on $R$ and $\nu$. Then, $h_\nu$ metric scale factor (8) can be written as

$$h_\nu = \frac{R}{\sqrt{1+\mu}}\frac{\partial W}{\partial \nu}\sqrt{S_\nu} \tag{A36}$$

in both the small-$\nu$ region and the large-$\nu$ region. That means the expression is formally the same for the small- and the large-$\nu$ regions.

With the help of (7), (A35), of Jensen's identity from [10] (particularly Eq. (52) with $t=1$) and of the relations for the generalized sine and cosine (14), (15)), the following identities can be derived in the small-$\nu$ region:

$$h_R^2 S_\nu = \left[\sum_{k=0}^{\infty}\binom{0-\mu k}{k}(W^2)^k\right]\left[\sum_{k=0}^{\infty}\binom{-(2+\mu)-\mu k}{k}(W^2)^k\right]$$

$$= \left[\sum_{k=0}^{\infty}\binom{0+(-1)-\mu k}{k}(W^2)^k\right]\left[\sum_{k=0}^{\infty}\binom{-(2+\mu)-(-1)-\mu k}{k}(W^2)^k\right]$$

$$= \left[\sum_{k=0}^{\infty}\binom{-1-\mu k}{k}(W^2)^k\right]\left[\sum_{k=0}^{\infty}\binom{-(1+\mu)-\mu k}{k}(W^2)^k\right] = f_C^2\frac{f_S^2}{W^2(1+\mu)} \tag{A37}$$

With the help of the same relations but using $t=1/(1+\mu)$ instead, we arrive in the large-$\nu$ region to the same relation as the one valid in the small-$\nu$ region. Then, the following relations in this sub-section are valid for both regions. When applying (A36) onto the left side of (A37), a very useful relation between the metric scale factors and the generalized sine and cosine arises:

$$h_R^2 h_\nu^2 (1+\mu)^2 = f_C^2 f_S^2 \frac{R^2}{W^2}\left(\frac{\partial W}{\partial \nu}\right)^2 \tag{A38}$$

Further, with a help of (16) and (17), we get from (A37)

$$h_R^2 S_\nu = \frac{(1+\mu)h_R^2 - 1}{\mu}\frac{\frac{1+\mu}{\mu}(1-h_R^2)}{W^2(1+\mu)} = \frac{1}{\mu^2 W^2}\left[-1+(2+\mu)h_R^2-(1+\mu)h_R^4\right] \tag{A39}$$

When we use the relation (A36) between $S_\nu$ and $h_\nu$, (A39) leads to the following relation between the metric scale factor $h_\nu$ and powers of $h_R$:

$$(1+\mu)h_\nu^2 = \frac{R^2}{\mu^2 W^2 \left(\frac{\partial W}{\partial \nu}\right)^2}\frac{1}{h_R^2}\left(-1+(2+\mu)h_R^2-(1+\mu)h_R^4\right) \tag{A40}$$

which is a very significant relation for the solution of the Laplace equation.

## A3 Derivatives of $W$ and $F(W)$ with respect to $\nu$ and $R$

### A3.1. Derivatives of W with respect to $\nu$ and $R$

The parameter $W$ (see Eq. (5)) and its derivatives (see Eqs. (9), (10)) are reported in the main text. On that basis, the second derivative with respect to $\nu$ is calculated as

$$\frac{\partial^2 W}{\partial \nu^2} = W\frac{\partial}{\partial \nu}\left(\frac{1+\mu\sin^2\nu}{\sin\nu\cos\nu}\right) + \frac{1+\mu\sin^2\nu}{\sin\nu\cos\nu}\frac{\partial W}{\partial \nu}$$

$$= \left(\frac{(1+\mu)\sin^2\nu-\cos^2\nu}{\sin^2\nu\cos^2\nu}\right)W + \left[\frac{1+\mu\sin^2\nu}{\sin\nu\cos\nu}\right]^2 W$$

$$= W\frac{(1+\mu)\sin^2\nu-\cos^2\nu+1+2\mu\sin^2\nu+\mu^2\sin^4\nu}{\sin^2\nu\cos^2\nu}$$

$$= \frac{2 + 3\mu + \mu^2 \sin^2 \nu}{\cos^2 \nu} W \tag{A41}$$

For completeness, the second derivative with respect to $R$ is

$$\frac{\partial^2 W}{\partial R^2} = \frac{\mu(\mu - 1)}{R^2} W \tag{A42}$$

A3.2. Derivatives of $F(W)$ with respect to $W$ and $\nu$

With a help of (9), the first derivative of a function $F$ depending on $W$ with respect to $\nu$ is

$$\frac{\partial F(W)}{\partial \nu} = \frac{dF(W)}{dW} \frac{\partial W}{\partial \nu} = \frac{dF(W)}{dW} \frac{1 + \mu \sin^2 \nu}{\sin \nu \cos \nu} W \tag{A43}$$

and the derivative with respect to $W$ is thus

$$\frac{dF(W)}{dW} = \frac{\partial F(W)}{\partial \nu} \bigg/ \frac{\partial W}{\partial \nu} = \frac{\partial F(W)}{\partial \nu} \frac{\sin \nu \cos \nu}{1 + \mu \sin^2 \nu} \frac{1}{W} \tag{A44}$$

The second derivative with respect to $W$ is

$$\begin{aligned}
\frac{\partial^2 F(W)}{\partial \nu^2} &= \frac{\partial}{\partial \nu} \left[ \frac{\partial F(W)}{\partial \nu} \right] = \frac{\partial}{\partial \nu} \left[ \frac{dF(W)}{dW} \frac{\partial W}{\partial \nu} \right] \\
&= \frac{\partial}{\partial \nu} \left[ \frac{dF(W)}{dW} \right] \frac{\partial W}{\partial \nu} + \frac{dF(W)}{dW} \frac{\partial}{\partial \nu} \left[ \frac{\partial W}{\partial \nu} \right] \\
&= \frac{d^2 F(W)}{dW^2} \left( \frac{\partial W}{\partial \nu} \right)^2 + \frac{\frac{\partial F(W)}{\partial \nu}}{\frac{\partial W}{\partial \nu}} \left[ \frac{\partial^2 W}{\partial \nu^2} \right]
\end{aligned} \tag{A45}$$

Then (with a help of (A41)),

$$\begin{aligned}
\frac{d^2 F(W)}{dW^2} &= \frac{\partial^2 F(W)}{\partial \nu^2} \frac{1}{\left(\frac{\partial W}{\partial \nu}\right)^2} - \frac{\partial F(W)}{\partial \nu} \frac{\left[\frac{\partial^2 W}{\partial \nu^2}\right]}{\left(\frac{\partial W}{\partial \nu}\right)^3} \\
&= \frac{\partial^2 F(W)}{\partial \nu^2} \frac{1}{\left(\frac{1+\mu \sin^2 \nu}{\sin \nu \cos \nu} W\right)^2} - \frac{\partial F(W)}{\partial \nu} \frac{\left[\frac{2+3\mu+\mu^2 \sin^2 \nu}{\cos^2 \nu} W\right]}{\left(\frac{1+\mu \sin^2 \nu}{\sin \nu \cos \nu} W\right)^3}
\end{aligned} \tag{A46}$$

Therefore, the second derivative of $F(W)$ with respect to $W$ is finally

$$\begin{aligned}
\frac{d^2 F(W)}{dW^2} &= \frac{\partial^2 F(W)}{\partial \nu^2} \frac{1}{W^2} \frac{\sin^2 \nu \cos^2 \nu}{(1 + \mu \sin^2 \nu)^2} \\
&- \frac{\partial F(W)}{\partial \nu} \frac{1}{W^2} \frac{\sin^3 \nu \cos \nu [2 + 3\mu + \mu^2 \sin^2 \nu]}{(1 + \mu \sin^2 \nu)^3}
\end{aligned} \tag{A47}$$

### A4 Derivatives of the Pólya-Szegő-type power series

An essential element of differential calculations in this article is the knowledge of the derivatives of the Pólya-Szegő-type power series present in the relations (A1), (A2) of this Appendix, with respect to $W$. These series, are, in what follows, restricted (see the conditions (A4) and (A5)) to the cases with $b = -\mu$, $\varepsilon = 2$, and with $b = \mu/(1 + \mu)$, $\varepsilon = -2/(1 + \mu)$, respectively, as these correspond to the series used in the small-$\nu$ region and in the large-$\nu$ region, respectively, in the SOS coordinate system. Therefore, they are particularly interesting for this article.

We define the following expressions and symbols for the two types of the series with generalized binomial coefficients:

$$S_C \equiv \sum_{k=0}^{\infty} \frac{a}{a - \mu k} \binom{a - \mu k}{k} (W^2)^k \text{ and } S_C \equiv W^{2a} \sum_{k=0}^{\infty} \frac{a}{a + bk} \binom{a + \frac{\mu}{1+\mu} k}{k} \left(W^{-\frac{2}{1+\mu}}\right)^k \tag{A48}$$

and

$$S_A \equiv \sum_{k=0}^{\infty} \binom{a - \mu k}{k} (W^2)^k \text{ and } S_A \equiv \frac{1}{1+\mu} W^{2a} \sum_{k=0}^{\infty} \binom{a + \frac{\mu}{1+\mu} k}{k} \left(W^{-\frac{2}{1+\mu}}\right)^k \tag{A49}$$

where the left side relations hold for the small-$\nu$ region, and the right side relations for the large-$\nu$ region. These expressions cover all the expressions with the power series employing generalized binomial coefficients which are used in the SOS coordinate system. Just $a$ parameter has to be set differently for the various elements of the system (relations for the coordinates transformation, the metric scale factors, the Jacobian, see [4], and the generalized sine and cosine, see [10]). Using (A48), (A49), (A7) and (7), it can be seen that

$$S_A = h_R^2 S_C \tag{A50}$$

in both regions.

A4.1. Derivative of $S_C$ type of Pólya-Szegő power series

We can write for both cases (i.e., for the small-$\nu$ region as well as for the large-$\nu$ region)

$$\frac{dS_C}{dW} = \frac{d}{dW}\left[W^\beta \sum_{k=0}^{\infty} \frac{a}{a+bk}\binom{a+bk}{k}(W^\varepsilon)^k\right] \tag{A51}$$

where $b = -\mu$, $\varepsilon = 2$, $\beta = 0$ in the small-$\nu$ region, and $b = \mu/(1+\mu)$, $\varepsilon = -2/(1+\mu)$, $\beta = 2a$ in the large-$\nu$ region. Then, the derivative can be relatively easily obtained by multiplication and division by a proper power of $W$ followed by the product rule use:

$$\begin{aligned}
\frac{dS_C}{dW} &= \frac{d}{dW}\left[W^\beta \sum_{k=0}^{\infty} \frac{a}{a+bk}\binom{a+bk}{k}(W^\varepsilon)^k\right] = \frac{d}{dW}\left[W^{\beta-\varepsilon\frac{a}{b}}W^{\varepsilon\frac{a}{b}}\sum_{k=0}^{\infty}\frac{a}{a+bk}\binom{a+bk}{k}(W^\varepsilon)^k\right] \\
&= \frac{d}{dW}\left[W^{\beta-\varepsilon\frac{a}{b}}\sum_{k=0}^{\infty}\frac{a}{a+bk}\binom{a+bk}{k}W^{\varepsilon k+\varepsilon\frac{a}{b}}\right] = \left(\beta-\varepsilon\frac{a}{b}\right)W^{\beta-\varepsilon\frac{a}{b}-1}\left[\sum_{k=0}^{\infty}\frac{a}{a+bk}\binom{a+bk}{k}W^{\varepsilon k+\varepsilon\frac{a}{b}}\right] \\
&+ W^{\beta-\varepsilon\frac{a}{b}}\left[\sum_{k=0}^{\infty}\frac{a}{a+bk}\binom{a+bk}{k}\left(\varepsilon k+\varepsilon\frac{a}{b}\right)W^{\varepsilon k+\varepsilon\frac{a}{b}-1}\right] = \left(\beta-\varepsilon\frac{a}{b}\right)W^{\beta-1}\left[\sum_{k=0}^{\infty}\frac{a}{a+bk}\binom{a+bk}{k}(W^\varepsilon)^k\right] \\
&+ W^{\beta-\varepsilon\frac{a}{b}}\varepsilon\frac{a}{b}\left[\sum_{k=0}^{\infty}\binom{a+bk}{k}\frac{bk+a}{a+bk}W^{\varepsilon k}W^{\varepsilon\frac{a}{b}-1}\right] = \left(\varepsilon\frac{a}{b}-\beta\right)W^{\beta-1}\left[-\sum_{k=0}^{\infty}\frac{a}{a+bk}\binom{a+bk}{k}(W^\varepsilon)^k\right] \\
&+ W^{\beta-1}\varepsilon\frac{a}{b}\left[\sum_{k=0}^{\infty}\binom{a+bk}{k}(W^\varepsilon)^k\right]
\end{aligned} \tag{A52}$$

Further simple algebraic manipulation leads to

$$\frac{dS_C}{dW} = \left(\varepsilon\frac{a}{b}-\beta\right)W^{\beta-1}\left\{-\left[\sum_{k=0}^{\infty}\frac{a}{a+bk}\binom{a+bk}{k}(W^\varepsilon)^k\right] + \frac{\varepsilon\frac{a}{b}}{\varepsilon\frac{a}{b}-\beta}\left[\sum_{k=0}^{\infty}\binom{a+bk}{k}(W^\varepsilon)^k\right]\right\} \tag{A53}$$

When the parameters for the small-$\nu$ region are input, we get

$$\frac{dS_C}{dW} = \frac{2a}{\mu W}\left\{\left[\sum_{k=0}^{\infty}\frac{a}{a-\mu k}\binom{a-\mu k}{k}(W^2)^k\right] - \left[\sum_{k=0}^{\infty}\binom{a-\mu k}{k}(W^2)^k\right]\right\} = \frac{2a}{\mu W}\{S_C - S_A\} \tag{A54}$$

By applying (A50) onto the second term, we get

$$\frac{dS_C}{dW} = \frac{2a}{\mu W}(1-h_R^2)S_C = \frac{2a}{W}\frac{f_S^2}{1+\mu}S_C \tag{A55}$$

in the small-$\nu$ region, where the rightmost expression comes from the relations between $f_S$ and $h_R$ (17).

When the parameters for the large-$\nu$ region are plugged into (A53), we get

$$\begin{aligned}
\frac{dS_C}{dW} &= \left(-\frac{2}{\mu}a - 2a\right)W^{2a-1}\left\{-\left[\sum_{k=0}^{\infty}\frac{a}{a+\frac{\mu}{1+\mu}k}\binom{a+\frac{\mu}{1+\mu}k}{k}\left(W^{-\frac{2}{1+\mu}}\right)^k\right]\right. \\
&\left.+\frac{-\frac{2}{\mu}a}{-\frac{2}{\mu}a-2a}\left[\sum_{k=0}^{\infty}\binom{a+\frac{\mu}{1+\mu}k}{k}\left(W^{-\frac{2}{1+\mu}}\right)^k\right]\right\} \\
&= \left(\frac{1}{\mu}+1\right)2aW^{-1}\left\{\left[W^{2a}\sum_{k=0}^{\infty}\frac{a}{a+\frac{\mu}{1+\mu}k}\binom{a+\frac{\mu}{1+\mu}k}{k}\left(W^{-\frac{2}{1+\mu}}\right)^k\right]\right.
\end{aligned}$$

$$-\frac{1}{1+\mu}\left[W^{2a}\sum_{k=0}^{\infty}\binom{a+\frac{\mu}{1+\mu}k}{k}\left(W^{-\frac{2}{1+\mu}}\right)^{k}\right]\Bigg\} = (1+\mu)\frac{2a}{\mu W}\{S_C - S_A\} \tag{A56}$$

By applying (A50) on the second term, we obtain

$$\frac{dS_C}{dW} = (1+\mu)\frac{2a}{\mu W}\{S_C - h_R^2 S_C\} = (1+\mu)\frac{2a}{\mu W}(1-h_R^2)S_C = \frac{2a}{W}f_S^2 S_C \tag{A57}$$

in the large-$\nu$ region, where the rightmost expression comes from the relations between $f_S$ and $h_R$ (17).

### A4.2. Derivative of $h_R^2$

More complicated is the derivative of the second-type power series $S_A$ (see Eq. (A49)). For this purpose, we will need to find first the derivative of $h_R^2$ with respect to $W$.

We can compare Eqs. (F19) and (F20) of Supplement F of [4] to determine a helpful identity applicable in the small-$\nu$ region

$$\sum_{k=1}^{\infty}\frac{k}{-1-\mu k}\binom{-1-\mu k}{k}(W^2)^k = W^2\sum_{k=0}^{\infty}\binom{-(\mu+2)-\mu k}{k}(W^2)^k \tag{A58}$$

Similarly, by comparison of Eqs. (F39) and (F40) of Supplement F of [10], another helpful identity applicable in the large-$\nu$ region is determined

$$\sum_{k=1}^{\infty}\frac{k}{-1+\frac{\mu}{1+\mu}k}\binom{-1+\frac{\mu}{1+\mu}k}{k}\left(W^{-\frac{2}{1+\mu}}\right)^{k} = W^{-\frac{2}{1+\mu}}\left[\sum_{k=0}^{\infty}\binom{-\frac{2+\mu}{1+\mu}+\frac{\mu}{1+\mu}k}{k}\left(W^{-\frac{2}{1+\mu}}\right)^{k}\right] \tag{A59}$$

Both these identities lead, in the respective regions, to the identity (see Eq. (F24) and (F44) of Supplement F of [4])

$$\frac{\partial h_R^2}{\partial \nu} = \frac{dh_R^2}{dW}\frac{\partial W}{\partial \nu} = -\frac{2\mu(1+\mu)}{R^2}\frac{W}{\frac{\partial W}{\partial \nu}}h_R^4 h_\nu^2 \tag{A60}$$

which is valid in both regions (small-$\nu$, large-$\nu$) when the proper series for the functions $h_R$, $h_\nu$ (see Eqs. (7), (8)) are used for the respective regions. Then (see Eqs. (8) and (A35))

$$\begin{aligned}\frac{dh_R^2}{dW} &= -\frac{2\mu(1+\mu)}{R^2}\frac{W}{\left(\frac{\partial W}{\partial \nu}\right)^2}h_R^4 h_\nu^2 \\ &= -\frac{2\mu(1+\mu)}{R^2}\frac{W}{\left(\frac{\partial W}{\partial \nu}\right)^2}h_R^4\frac{R^2}{1+\mu}\left(\frac{\partial W}{\partial \nu}\right)^2\sum_{k=0}^{\infty}\binom{-(\mu+2)-\mu k}{k}(W^2)^k \\ &= -2\mu W h_R^4 \sum_{k=0}^{\infty}\binom{-(\mu+2)-\mu k}{k}(W^2)^k = -2\mu W h_R^4 S_\nu\end{aligned} \tag{A61}$$

in the small-$\nu$ region, while

$$\begin{aligned}\frac{dh_R^2}{dW} &= -\frac{2\mu(1+\mu)}{R^2}\frac{W}{\left(\frac{\partial W}{\partial \nu}\right)^2}h_R^4 h_\nu^2 = -\frac{2\mu(1+\mu)}{R^2}\frac{W}{\left(\frac{\partial W}{\partial \nu}\right)^2}h_R^4\frac{R^2}{(1+\mu)^2}W^{-2\frac{2+\mu}{1+\mu}}\left(\frac{\partial W}{\partial \nu}\right)^2\sum_{k=0}^{\infty}\binom{-\frac{2+\mu}{1+\mu}+\frac{\mu}{1+\mu}k}{k}\left(W^{-\frac{2}{1+\mu}}\right)^{k} \\ &= -\frac{2\mu}{1+\mu}h_R^4 W W^{-2-\frac{2}{1+\mu}}\sum_{k=0}^{\infty}\binom{-\frac{2+\mu}{1+\mu}+\frac{\mu}{1+\mu}k}{k}\left(W^{-\frac{2}{1+\mu}}\right)^{k} = -2\mu W h_R^4 S_\nu\end{aligned} \tag{A62}$$

in the large-$\nu$ region. A correct relation for $S_\nu$ according to (A35) has to be used for the particular region. Then, the formal expression for the derivative is the same as in the small-$\nu$ region.

The final formulae in (A61), (A62) have an advantage with respect to the initial formulae that they do not contain explicitly $R$, which is a correct result (thanks to the $h_\nu$ dependence on $R$). The given formulae for the derivative $\frac{dh_R^2}{dW}$ thus does not depend explicitly on $R$ (it depends directly only on $W$). (The use of $h_\nu$ symbol and division by $R^2$—as is the case in the initial formula—would be rather misleading).

### A4.3. Derivative of $S_A$ type of Pólya-Szegő power series

Now, we have the tools for derivation of the $S_A$ power series with respect to $W$ in terms of the already known functions.

Thanks to the identity (A50), we can write

$$\frac{dS_A}{dW} = \frac{d}{dW}\left[h_R^2 S_C\right] = \frac{dh_R^2}{dW}S_C + h_R^2\frac{dS_C}{dW} \tag{A63}$$

Using (A61) and (A55),

$$\frac{dS_A}{dW} = -2\mu W h_R^4 S_\nu S_C + h_R^2 \frac{2a}{\mu W}(1 - h_R^2) S_C \tag{A64}$$

in the small-$\nu$ region. Further, we again use the identity (A50) and obtain

$$\frac{dS_A}{dW} = -2\mu W h_R^2 S_\nu S_A + \frac{2a}{\mu W}(1 - h_R^2) S_A \tag{A65}$$

in the small-$\nu$ region. As further the relation (A39) for $h_R^2 S_\nu$ holds, we obtain

$$\begin{aligned}\frac{dS_A}{dW} &= \left[\frac{2a}{\mu W}(1 - h_R^2) - 2\frac{1}{\mu W}(-1 + (2 + \mu)h_R^2 - (1 + \mu)h_R^4)\right] S_A \\ &= \frac{2}{\mu W}\left[a(1 - h_R^2) - (-1 + (2 + \mu)h_R^2 - (1 + \mu)h_R^4)\right] S_A \\ &= \frac{2}{\mu W}\left[(a + 1) - (2 + \mu + a)h_R^2 + (1 + \mu)h_R^4\right] S_A\end{aligned} \tag{A66}$$

Further form of the formula—the one containing generalized sinus $f_S$—can be derived with a help of the expression (17):

$$\frac{dS_A}{dW} = \frac{2}{W(1 + \mu)}\left[\mu f_S^2 + (a - \mu)\right] f_S^2 S_A \tag{A67}$$

These formulas are valid in the small-$\nu$ region.

Using (A62) and (A57),

$$\frac{dS_A}{dW} = -2\mu W h_R^4 S_\nu S_C + h_R^2(1 + \mu)\frac{2a}{\mu W}(1 - h_R^2) S_C \tag{A68}$$

in the large-$\nu$ region. When we use again the identity (A50) onto (A68) in the large-$\nu$ region, we get

$$\frac{dS_A}{dW} = -2\mu W h_R^2 S_\nu S_A + (1 + \mu)\frac{2a}{\mu W}(1 - h_R^2) S_A \tag{A69}$$

As further the relation (A39) for $h_R^2 S_\nu$ holds, we obtain

$$\begin{aligned}\frac{dS_A}{dW} &= \left[(1 + \mu)\frac{2a}{\mu W}(1 - h_R^2) - 2\frac{1}{\mu W}(-1 + (2 + \mu)h_R^2 - (1 + \mu)h_R^4)\right] S_A \\ &= \frac{2}{\mu W}\left[(1 + \mu)a(1 - h_R^2) - (-1 + (2 + \mu)h_R^2 - (1 + \mu)h_R^4)\right] S_A \\ &= \frac{2}{\mu W}\left[(1 + \mu)a - (1 + \mu)a h_R^2 + 1 - (2 + \mu)h_R^2 + (1 + \mu)h_R^4\right] S_A \\ &= \frac{2}{\mu W}\left[(1 + \mu)a + 1 - (2 + \mu + (1 + \mu)a)h_R^2 + (1 + \mu)h_R^4\right] S_A\end{aligned} \tag{A70}$$

Further form of the formula—the one containing generalized sinus $f_S$—can be derived with a help of the expression (17):

$$\frac{dS_A}{dW} = \frac{2}{W(1 + \mu)}\left[\mu f_S^2 + ((1 + \mu)a - \mu)\right] f_S^2 S_A \tag{A71}$$

Formulas (A70) and (A71) are valid in the large-$\nu$ region.

A4.4. Summary for the derivative of $S_C$ and $S_A$ type of Pólya-Szegő power series

By the above analysis, we derived the derivatives of the expressions $S_C$ and $S_A$, which frequently appear with a varying parameter $a$ in the SOS coordinates expressions, for both the in the small- and the large-$\nu$ regions. The derivatives are expressed with a help of the already known power series: $S_C$ or $S_A$ as well as $h_R$ or $f_S$. Notice that the derivatives of $S_C$ and $S_A$ in both regions are proportional to the function itself.

The derivatives are derived with respect to $W$. The derivatives with respect to $\nu$ or $R$ can be obtained by multiplying the above formulas by $\frac{\partial W}{\partial \nu}$ or $\frac{\partial W}{\partial R}$ which can be found in the relations (9), (10).

Further notice that the expressions for $S_C$ and $S_A$ derivatives are formally different in the small- and in the large-$\nu$ region. Nevertheless, also notice that the parameter $a$ is differing (for the power series in the same function type, e.g., in the $h_\nu$ metric scale factor or in the Jacobian) in the two regions as well. For the large-$\nu$ region, the parameter $a$ is $(1 + \mu)$-times smaller than in the small-$\nu$ region. Therefore, the physically relevant expressions derived in what follows will be formally written in the same form in both regions.

A4.5. Derivatives of power series contained in the metric scale factors

For special values of the parameter $a$ in the derivatives of $S_A$ expressions, we get derivatives of the metric scale factors.

The first interim relation provides derivative of $S_\nu$ (see Eq. (A35), connected with $h_\nu$ metric scale factor by (A36)). It can be calculated from the $S_A$ derivative (see Eqs. (A65), (A69)) taking into account that $S_\nu = S_A$ for $a = -(\mu + 2)$ in the small-$\nu$ region, and for $a = -(\mu + 2)/(1 + \mu)$ in the large-$\nu$ region. Input of the parameter $a$ for the small-$\nu$ region leads (using (A65)) to

$$\frac{dS_\nu}{dW} = \frac{dS_A}{dW} = \frac{\partial \sum_{k=0}^\infty \binom{-(\mu+2)-\mu k}{k}(W^2)^k}{\partial W} = \left[\frac{-2(\mu+2)}{\mu W}(1-h_R^2) - 2\mu W h_R^2 S_\nu\right] S_\nu \tag{A72}$$

The same action in the large-$\nu$ region leads—using (A69)—to

$$\begin{aligned}\frac{dS_\nu}{dW} = \frac{dS_A}{dW} &= -2\mu W h_R^2 S_\nu S_\nu + (1+\mu)\frac{-2\frac{\mu+2}{1+\mu}}{\mu W}(1-h_R^2)S_\nu \\ &= \left[-2\mu W h_R^2 S_\nu + \frac{-2(\mu+2)}{\mu W}(1-h_R^2)\right]S_\nu\end{aligned} \tag{A73}$$

As can be seen, the relations are the same for the small and the large-$\nu$ regions when proper relations for $S_\nu$ and $h_R$ are used in the respective region.

The second set of the relations should provide derivative of $h_R$. Nevertheless, these were already derived above in (A61) and (A62) with the following result:

$$\frac{dh_R^2}{dW} = -2\mu W h_R^4 S_\nu \tag{A74}$$

Again, these relations are the same for the small- and the large-$\nu$ region when the proper relations for $S_\nu$ and $h_R$ are used in the respective region. With a help of (A39), it can be also written in the following forms:

$$\begin{aligned}\frac{dh_R^2}{dW} &= \frac{2}{\mu W}\left(h_R^2 - (2+\mu)h_R^4 + (1+\mu)h_R^6\right) \\ &= -\frac{2}{\mu W}h_R^2(1-h_R^2)\left[(1+\mu)h_R^2 - 1\right] = -\frac{\mu}{1+\mu}\frac{2}{W}h_R^2 f_S^2 f_C^2\end{aligned} \tag{A75}$$

valid also in both the small- and the large-$\nu$ region. The last expression is obtained with the help of the relation (A22). Further,

$$\frac{dh_R}{dW} = \frac{1}{2h_R}\frac{dh_R^2}{dW} = \mu W h_R^3 S_\nu = -\frac{h_R}{\mu W}\left(-1 + (2+\mu)h_R^2 - (1+\mu)h_R^4\right) = -\frac{\mu}{1+\mu}\frac{h_R}{W}f_S^2 f_C^2 \tag{A76}$$

More generally, we have

$$\frac{d(h_R)^j}{dW} = j(h_R)^{j-1}\frac{dh_R}{dW} = -j\frac{(h_R)^j}{\mu W}\left(-1 + (2+\mu)h_R^2 - (1+\mu)h_R^4\right) = -j\frac{\mu}{1+\mu}\frac{(h_R)^j}{W}f_S^2 f_C^2 \tag{A77}$$

valid for any integer $j$.

A4.6. Derivatives of the generalized cosine and sine

We can also derive the derivatives of the squared generalized cosine $f_C$ and sine $f_S$. With a help of (14) and (A66), $S_A$ according to the definition (A49) and with $a = -1$, we get in the small-$\nu$ region

$$\begin{aligned}\frac{df_C^2}{dW} &= \frac{d}{dW}\left[\sum_{k=0}^\infty \binom{-1-\mu k}{k}(W^2)^k\right] = \frac{dS_A}{dW} \\ &= \frac{2}{\mu W}\left[(-1+1) - (2+\mu-1)h_R^2 + (1+\mu)h_R^4\right]S_A \\ &= -\frac{2}{\mu W}(1+\mu)h_R^2\left[1 - h_R^2\right]f_C^2 = -\frac{2}{W}h_R^2 f_S^2 f_C^2 = \frac{1+\mu}{\mu}\frac{dh_R^2}{dW}\end{aligned} \tag{A78}$$

where the last formula follows from algebraic manipulation using identities (A61) and (A37). In the large-$\nu$ region, by comparing (14) and (A49) with $a = -1/(1+\mu)$), and by using (A70),

$$\begin{aligned}\frac{df_C^2}{dW} &= \frac{d}{dW}\left[\frac{W^{-\frac{2}{1+\mu}}}{1+\mu}\sum_{k=0}^{\infty}\binom{-\frac{1}{1+\mu}+\frac{\mu}{1+\mu}k}{k}\left(W^{-\frac{2}{1+\mu}}\right)^k\right]\\ &= \frac{dS_A}{dW} = \frac{2}{\mu W}\left[(1+\mu)\frac{-1}{1+\mu}+1-\left(2+\mu+(1+\mu)\frac{-1}{1+\mu}\right)h_R^2+(1+\mu)h_R^4\right]S_A\\ &= \frac{2}{\mu W}[-(1+\mu)h_R^2+(1+\mu)h_R^4]S_A = \frac{2}{\mu W}h_R^2(1+\mu)[-1+h_R^2]f_C^2\\ &= -\frac{2}{\mu W}(1+\mu)h_R^2[1-h_R^2]f_C^2 = -\frac{2}{W}h_R^2 f_S^2 f_C^2 = \frac{1+\mu}{\mu}\frac{dh_R^2}{dW}\end{aligned} \qquad (A79)$$

The relations are formally the same in both regions.

Thanks to the (16) relation, the derivative of the square of the second generalized trigonometric function, $f_S$, with respect to $W$ is equal but of the opposite sign than the derivative of the squared generalized cosine $f_C$:

$$\frac{df_S^2}{dW} = \frac{d[1-f_C^2]}{dW} = \frac{2}{W}h_R^2 f_S^2 f_C^2 = -\frac{1+\mu}{\mu}\frac{dh_R^2}{dW} \qquad (A80)$$

Also for this generalized trigonometric function, the relation is valid for both the small- and the large-$\nu$ region. This simplifies all further derivations containing derivatives of the generalized sine and cosine, which can be done with the same notation in both regions.

It should be noted that the above derivatives are carried out with respect to the parameter $W$, not with respect to the coordinate $\nu$. It means that when we want to compare limiting values of derivatives of generalized trigonometric functions with the derivatives of the normal trigonometric functions, we have to calculate the derivative with respect to $\nu$ first, and only afterward to set $\mu = 0$. The chain rule tells that

$$\frac{\partial f_S^2}{\partial \nu} = \frac{df_S^2}{dW}\frac{\partial W}{\partial \nu} = \frac{2}{W}h_R^2 f_S^2 f_C^2 \frac{1+\mu\sin^2\nu}{\sin\nu\cos\nu}W = 2h_R^2 f_S^2 f_C^2\frac{1+\mu\sin^2\nu}{\sin\nu\cos\nu} \qquad (A81)$$

where (9) was used for the partial derivative. For the limiting value $\mu = 0$ corresponding to the spherical case, we have (as $h_R = 1$, $f_C$ being equal to cosine, $f_S$ being equal to sine in such case)

$$\frac{\partial f_S^2}{\partial \nu} = 2.1.\sin^2\nu\cos^2\nu\frac{1+0.\sin^2\nu}{\sin\nu\cos\nu} = 2\sin\nu\cos\nu \qquad (A82)$$

which is equal to the derivative of the squared sine with respect to $\nu$.

The derivatives of the generalized cosine and sine (i.e., not squared) are

$$\frac{df_C}{dW} = \frac{1}{2f_C}\frac{df_C^2}{dW} = -\frac{1}{W}h_R^2 f_S^2 f_C \qquad (A83)$$

and

$$\frac{df_S}{dW} = \frac{1}{W}h_R^2 f_C^2 f_S \qquad (A84)$$

Still more generally, the derivatives of the powers of the generalized sine and cosine are

$$\frac{d[(f_C)^m]}{dW} = m(f_C)^{m-1}\frac{df_C}{dW} = -m(f_C)^{m-1}\frac{1}{W}h_R^2 f_S^2 f_C = -mf_C^m\frac{1}{W}h_R^2 f_S^2 \qquad (A85)$$

$$\frac{d[(f_S)^m]}{dW} = m(f_S)^{m-1}\frac{df_S}{dW} = m(f_S)^{m-1}\frac{1}{W}h_R^2 f_C^2 f_S = mf_S^m\frac{1}{W}h_R^2 f_C^2 \qquad (A86)$$

Other important derivatives of the special composed functions containing generalized sine or cosine are as follows:

$$\frac{d}{dW}\left[\frac{f_C^2}{W}\right] = \frac{d}{dW}\left[\frac{1}{W}\right]f_C^2 + \frac{1}{W}\frac{d}{dW}[f_C^2] = -\frac{1}{W^2}f_C^2 - \frac{1}{W}\frac{2}{W}h_R^2 f_S^2 f_C^2 = -\frac{f_C^2}{W^2}(2h_R^2 f_S^2 + 1) \qquad (A87)$$

$$\frac{d}{dW}\left[\frac{f_S^2}{W}\right] = \frac{d}{dW}\left[\frac{1}{W}\right]f_S^2 + \frac{1}{W}\frac{d}{dW}[f_S^2] = -\frac{1}{W^2}f_S^2 + \frac{1}{W}\frac{2}{W}h_R^2 f_S^2 f_C^2 = \frac{f_S^2}{W^2}(2h_R^2 f_C^2 - 1) \qquad (A88)$$

Further useful derivative is the derivative of the ratio of generalized sine and cosine:

$$\frac{d\left(\frac{f_S}{f_C}\right)}{dW} = \frac{\frac{d(f_S)}{dW}f_C - f_S\frac{d(f_C)}{dW}}{f_C^2} = \frac{h_R^2}{W}\left(f_C f_S + \frac{1}{f_C}f_S^3\right) = \frac{h_R^2}{W}\frac{f_S}{f_C}(f_C^2 + f_S^2) = \frac{h_R^2}{W}\frac{f_S}{f_C} \tag{A89}$$

and also derivative of its square:

$$\frac{d\left[\left(\frac{f_S}{f_C}\right)^2\right]}{dW} = 2\left(\frac{f_S}{f_C}\right)\frac{d\left(\frac{f_S}{f_C}\right)}{dW} = 2\left(\frac{f_S}{f_C}\right)\frac{h_R^2}{W}\frac{f_S}{f_C} = 2\frac{h_R^2}{W}\left(\frac{f_S}{f_C}\right)^2 \tag{A90}$$

The reversed ratio of generalized goniometric functions has derivative

$$\frac{d\left(\frac{f_C}{f_S}\right)}{dW} = -\frac{h_R^2}{W}\left(f_C f_S + \frac{1}{f_S}f_C^3\right) = -\frac{h_R^2}{W}\frac{f_C}{f_S}(f_S^2 + f_C^2) = -\frac{h_R^2}{W}\frac{f_C}{f_S} \tag{A91}$$

Finally, a very important derivative is the one of the ratio of the generalized sine $f_S$ and the metric scale factor $h_R$. First, let us find the derivative of the square of it. With the help of (17) and (A75), we easily obtain

$$\frac{d\left(\frac{f_S^2}{h_R^2}\right)}{dW} = \frac{\frac{df_S^2}{dW}h_R^2 - f_S^2\frac{dh_R^2}{dW}}{h_R^4} = \frac{-\frac{1+\mu}{\mu}\frac{dh_R^2}{dW}h_R^2 - \frac{1+\mu}{\mu}(1-h_R^2)\frac{dh_R^2}{dW}}{h_R^4} = \frac{-\frac{1+\mu}{\mu}\frac{dh_R^2}{dW}}{h_R^4} = \frac{2}{W}f_C^2\frac{f_S^2}{h_R^2} \tag{A92}$$

Thus

$$\frac{d\left(\frac{f_S}{h_R}\right)}{dW} = \frac{1}{2\left(\frac{f_S}{h_R}\right)}\frac{d\left(\frac{f_S^2}{h_R^2}\right)}{dW} = \frac{f_C^2}{W}\frac{f_S}{h_R} \tag{A93}$$

and generally (also with the help of (17))

$$\frac{d\left[\left(\frac{f_S}{h_R}\right)^m\right]}{dW} = m\frac{f_C^2}{W}\left(\frac{f_S}{h_R}\right)^m = m\frac{(1+\mu)h_R^2 - 1}{\mu W}\left(\frac{f_S}{h_R}\right)^m \tag{A94}$$

A remarkable property of $(f_S/h_R)^m$ derivative is that the derivative (and the second derivative) gives the same function $(f_S/h_R)^m$ multiplied by $m$ and $m$-independent function. Remind also that the above derivatives are with respect to $W$, not $\nu$.

The power $m$ generalized cosine over metric scale factor derivative can be derived accordingly:

$$\frac{d\left[\left(\frac{f_C}{h_R}\right)^m\right]}{dW} = -m\frac{f_S^2}{(1+\mu)W}\left(\frac{f_C}{h_R}\right)^m \tag{A95}$$

A4.7. Derivatives of functions containing Jacobian

For the Laplace equation solution, several derivatives of the expressions containing Jacobian are needed. First, we derive the derivative of the Jacobian divided by the square of the $h_R$ divided by $\frac{\partial W}{\partial \nu}$ for the small-$\nu$ region with respect to $W$. With the help of (12), and using (A55), i.e., the derivative of $S_C$ type of power series, we can derive the following:

$$\begin{aligned}\frac{\partial}{\partial R}\left(\frac{\Im}{h_R^2\frac{\partial W}{\partial \nu}}\right) &= \frac{\partial}{\partial R}\left[\frac{R^2}{\sqrt{1+\mu}}\sum_{k=0}^{\infty}\frac{-\frac{\mu+3}{2}}{-\frac{\mu+3}{2}-\mu k}\binom{-\frac{\mu+3}{2}-\mu k}{k}(W^2)^k\right] \\ &= \frac{\partial}{\partial R}\left[\frac{R^2}{\sqrt{1+\mu}}S_C\right] = \frac{2R}{\sqrt{1+\mu}}S_C + \frac{R^2}{\sqrt{1+\mu}}\frac{dS_C}{dW}\frac{\partial W}{\partial R} = \frac{R}{\sqrt{1+\mu}}\left(2S_C + R\frac{dS_C}{dW}\frac{\mu}{R}W\right) \\ &= \frac{R}{\sqrt{1+\mu}}\left[2S_C + R\frac{2\left(-\frac{\mu+3}{2}\right)}{\mu W}(1-h_R^2)S_C\frac{\mu}{R}W\right] = \frac{R}{\sqrt{1+\mu}}[2-(\mu+3)(1-h_R^2)]S_C \\ &= \frac{R}{\sqrt{1+\mu}}[-(1+\mu)+(\mu+3)h_R^2]\frac{\sqrt{1+\mu}}{R^2}\frac{\Im}{h_R^2\frac{\partial W}{\partial \nu}} = \frac{\Im}{R\frac{\partial W}{\partial \nu}}\left[(\mu+3) - \frac{1+\mu}{h_R^2}\right] \end{aligned} \tag{A96}$$

Further, we will need the derivative of the Jacobian divided by the square of the $h_\nu$, multiplied by $\frac{\partial W}{\partial \nu}$ for the small-$\nu$ region with respect to $W$. With the help of (13) and using (A55), we arrive at

$$\frac{\partial}{\partial W}\left(\frac{\Im}{h_\nu^2}\frac{\partial W}{\partial \nu}\right) = \frac{\partial}{\partial W}\left(\sqrt{1+\mu}\sum_{k=0}^{\infty}\frac{\frac{\mu+1}{2}}{\frac{\mu+1}{2}-\mu k}\binom{\frac{\mu+1}{2}-\mu k}{k}(W^2)^k\right)$$
$$= \frac{\partial}{\partial W}\left[\sqrt{1+\mu}S_C\right] = \sqrt{1+\mu}\frac{dS_C}{dW} = \sqrt{1+\mu}\left(\frac{2\left(\frac{\mu+1}{2}\right)}{\mu W}(1-h_R^2)S_C\right) \quad \text{(A97)}$$
$$= \sqrt{1+\mu}\left(\frac{(\mu+1)}{\mu W}(1-h_R^2)\frac{1}{\sqrt{1+\mu}}\frac{\Im}{h_\nu^2}\frac{\partial W}{\partial \nu}\right) = \frac{1+\mu}{\mu W}(1-h_R^2)\frac{\Im}{h_\nu^2}\frac{\partial W}{\partial \nu}$$

The above derivative with respect to the coordinate $\nu$ is

$$\frac{\partial}{\partial \nu}\left(\frac{\Im}{h_\nu^2}\frac{\partial W}{\partial \nu}\right) = \frac{\partial}{\partial W}\left(\frac{\Im}{h_\nu^2}\frac{\partial W}{\partial \nu}\right)\frac{\partial W}{\partial \nu} = \frac{1+\mu}{\mu W}(1-h_R^2)\frac{\Im}{h_\nu^2}\frac{\partial W}{\partial \nu}\frac{\partial W}{\partial \nu} = \frac{f_S^2}{W}\frac{\Im}{h_\nu^2}\left(\frac{\partial W}{\partial \nu}\right)^2 \quad \text{(A98)}$$

### A5 Integral identities involving the Pólya-Szegő-type power series

The integral

$$\int W h_R^2 dW = \int W \sum_{k=0}^{\infty}\binom{-\mu k}{k}(W^2)^k dW \quad \text{(A99)}$$

can be expressed in the following form:

$$\int W h_R^2 dW = \int W \sum_{k=0}^{\infty}\binom{-\mu k}{k}\left(\left(W^{\frac{1}{\mu}}\right)^{2\mu}\right)^k \frac{W\mu}{W^{\frac{1}{\mu}}}dW^{\frac{1}{\mu}}$$
$$= \int \sum_{k=0}^{\infty}\binom{-\mu k}{k}\left(W^{\frac{1}{\mu}}\right)^{2\mu k+2\mu-1}\mu dW^{\frac{1}{\mu}}$$
$$= \mu\sum_{k=0}^{\infty}\binom{-\mu k}{k}\frac{1}{2\mu k+2\mu}\left(W^{\frac{1}{\mu}}\right)^{2\mu k+2\mu} + \text{const.}$$
$$= \frac{1}{2}\sum_{k=0}^{\infty}\binom{-\mu k}{k}\frac{1}{k+1}(W^2)^{k+1} + \text{const.} \quad \text{(A100)}$$

Now, we can employ the basic binomial identity (see [12])

$$\binom{\alpha}{k} = \frac{\alpha}{k}\binom{\alpha-1}{k-1} \Rightarrow \binom{\alpha}{k}\frac{\alpha+1}{k+1} = \binom{\alpha+1}{k+1} \quad \text{(A101)}$$

and arrive at (we skip for the moment the integration constant and attach it again at the end of derivation only)

$$\int W h_R^2 dW = \frac{1}{2}\sum_{k=0}^{\infty}\frac{1}{-\mu k+1}\binom{-\mu k}{k}\frac{-\mu k+1}{k+1}(W^2)^{k+1}$$
$$= \frac{1}{2}\sum_{k=0}^{\infty}\frac{1}{-\mu k+1}\binom{-\mu k+1}{k+1}(W^2)^{k+1}$$
$$= \frac{1}{2}\sum_{M=1}^{\infty}\frac{1}{-\mu(M-1)+1}\binom{-\mu(M-1)+1}{(M-1)+1}(W^2)^{(M-1)+1}$$
$$= \frac{1}{2}\sum_{M=1}^{\infty}\frac{1}{(1+\mu)-\mu M}\binom{(1+\mu)-\mu M}{M}(W^2)^M$$
$$= \frac{1}{2}\sum_{M=1}^{\infty}\frac{1}{(1+\mu)-\mu M}\binom{(1+\mu)-\mu M}{M}(W^2)^M\frac{1+\mu}{1+\mu} + \frac{1}{2(1+\mu)} - \frac{1}{2(1+\mu)}$$
$$= \frac{1}{2(1+\mu)}\left[\sum_{M=0}^{\infty}\frac{1+\mu}{(1+\mu)-\mu M}\binom{(1+\mu)-\mu M}{M}(W^2)^M - 1\right] \quad \text{(A102)}$$

With the help of the expression (A7) and (15), we can express it as follows:

$$\int W h_R^2 dW = \frac{1}{2(1+\mu)} \left[ \frac{\sum_{k=0}^{\infty} \binom{-\mu k}{k}(W^2)^k}{\sum_{k=0}^{\infty} \binom{-(1+\mu)-\mu k}{k}(W^2)^k} - 1 \right] = \frac{1}{2(1+\mu)} \left[ \frac{h_R^2}{f_S^2} W^2 (1+\mu) - 1 \right] \quad (A103)$$

Then, the integral can be expressed in the following form:

$$\int W h_R^2 dW = \frac{W^2}{2} \frac{h_R^2}{f_S^2} - \frac{1}{2(1+\mu)} + const. \quad (A104)$$

Another very important integral is $\int W^{-1} h_R^2 dW$. It can be derived in two ways. The first one is a direct integration of the infinite power series. In the small-$\nu$ region, it is

$$\int W^{-1} h_R^2 dW = \int W^{-1} \sum_{k=0}^{\infty} \binom{-\mu k}{k}(W^2)^k dW \quad (A105)$$

and thus

$$\int W^{-1} h_R^2 dW = \int \left[ \frac{1}{W} + \sum_{k=1}^{\infty} \binom{-\mu k}{k}(W)^{2k-1} \right] dW$$
$$= \ln W + \sum_{k=1}^{\infty} \binom{-\mu k}{k} \frac{1}{2k}(W)^{2k} + const. \quad (A106)$$
$$= \ln W + \frac{1}{2} \sum_{k=1}^{\infty} \binom{-\mu k}{k} \frac{1}{k}(W^2)^k + const.$$

Alternatively, we can also employ (A75) relation for the derivative of the squared $h_R$, from which we see that

$$\frac{dW}{\mu W} = \frac{d(h_R^2)}{2 h_R^2 (1 - (2+\mu) h_R^2 + (1+\mu) h_R^4)} \Rightarrow \frac{h_R^2}{W} dW$$
$$= \frac{\mu}{2(1 - (2+\mu) h_R^2 + (1+\mu) h_R^4)} d(h_R^2) \quad (A107)$$

Therefore,

$$\int W^{-1} h_R^2 dW = \frac{1}{2} \int \frac{\mu}{1 - (2+\mu) h_R^2 + (1+\mu) h_R^4} d(h_R^2) \quad (A108)$$

This integral can be found in an analytical form:

$$\int W^{-1} h_R^2 dW = \frac{\mu}{2} \left\{ \frac{i \ln\left[1 - \frac{i(2(1+\mu) h_R^2 - (2+\mu))}{\sqrt{-\mu^2}}\right]}{\sqrt{-\mu^2}} - \frac{i \ln\left[1 + \frac{i(2(1+\mu) h_R^2 - (2+\mu))}{\sqrt{-\mu^2}}\right]}{\sqrt{-\mu^2}} \right\}$$
$$= \frac{\mu}{2|\mu|} \left\{ \ln\left[1 - \frac{(2(1+\mu) h_R^2 - (2+\mu))}{|\mu|}\right] - \ln\left[1 + \frac{(2(1+\mu) h_R^2 - (2+\mu))}{|\mu|}\right] \right\} \quad (A109)$$

For oblate spheroids, $\mu$ is positive, and thus

$$\int W^{-1} h_R^2 dW = \frac{1}{2} \left\{ \ln\left[\frac{2(1+\mu)(1-h_R^2)}{\mu}\right] - \ln\left[\frac{2((1+\mu) h_R^2 - 1)}{\mu}\right] \right\} = \frac{1}{2} \{\ln[2 f_S^2] - \ln[2 f_C^2]\} \quad (A110)$$

where we used the relations (17) between $h_R$ and the generalized cosine and sine. Thus,

$$\int W^{-1} h_R^2 dW = \frac{1}{2} \ln\left[\frac{f_S^2}{f_C^2}\right] + const. = \ln\left[\frac{f_S}{f_C}\right] + const. \quad (A111)$$

The resulting relation is in fact logarithm of the generalized tangent. The correctness of the derivation can be easily checked with a help of the derivatives of the generalized cosine and sine (A78) and (A80). Note that a comparison of both alternatives of derivation of this integral, (A106) and (A111), leads to a new binomial identity:

$$\sum_{k=1}^{\infty} \binom{-\mu k}{k} \frac{1}{k}(W^2)^k = \ln\left[\left(\frac{f_S}{W f_C}\right)^2\right] = \ln\left[\frac{1}{W^2} \frac{(1+\mu)(1-h_R^2)}{(1+\mu) h_R^2 - 1}\right] \quad (A112)$$

where the final algebraic manipulation is based on the relation (17) between $h_R$ and the generalized cosine and sine.